\documentclass[aps,prd,showpacs,floats,onecolumn,nofootinbib]{revtex4}
\pdfoutput=1  
\usepackage{graphicx,color}
\usepackage{subfigure}
\usepackage{latexsym,amsmath,amssymb,graphicx,booktabs}
\usepackage{hyperref}

\definecolor{MyBlue}{rgb}{0.15,0.15,0.70}

\hypersetup{
colorlinks=true,
citecolor=MyBlue,
linkcolor=MyBlue,
urlcolor=MyBlue
}

\usepackage{amssymb}
\usepackage{amsmath}
\usepackage{amsfonts}
\usepackage{upgreek}
\usepackage{latexsym}

\newcommand{\bk}{{\mathbf k}}
\newcommand{\bfe}{{\mathbf e}}

\newcommand{\HH}{{\cal H}}

\newcommand{\LL}{{\cal L}}

\newcommand{\al}{\alpha}

\newcommand{\de}{\delta}

\newcommand{\ep}{\epsilon}
\newcommand{\ga}{\gamma}

\newcommand{\La}{\Lambda}

\newcommand{\gsim}{\stackrel{>}{\sim}}
\newcommand{\lsim}{\stackrel{<}{\sim}}
\newcommand{\bea}{\begin{eqnarray}}
\newcommand{\eea}{\end{eqnarray}}
\newcommand{\bean}{\begin{eqnarray*}}
\newcommand{\eean}{\end{eqnarray*}}

\def\id{{\rm 1\kern -2.5pt I}}

\graphicspath{ {./Graphs3/} }

\newcommand{\ra}{\rightarrow}

\def\lsim{\raise 0.4ex\hbox{$<$}\kern -0.8em\lower 0.62
ex\hbox{$\sim$}}

\def\gsim{\raise 0.4ex\hbox{$>$}\kern -0.7em\lower 0.62
ex\hbox{$\sim$}}

\def\lbar{{\hbox{$\lambda$}\kern -0.7em\raise 0.6ex
\hbox{$-$}}}

\newcommand\p{\partial}

\newcommand\ee{\end{equation}}
\newcommand\be{\begin{equation}}
\def\bea{\begin{array}}
\def\eea{\end{array}}\def\ea{\end{array}}
\newcommand\ees{\end{eqnarray}}
\newcommand\bees{\begin{eqnarray}}
\def\nn{\nonumber}

\newcommand{\om}{\omega}
\newcommand{\Om}{\Omega}

\def\dslash{\hspace{-1mm}\not{\hbox{\kern-2pt $\partial$}}}
\def\Dslash{\not{\hbox{\kern-4pt $D$}}}
\def\pslash{\not{\hbox{\kern-2.1pt $p$}}}
\def\kslash{\not{\hbox{\kern-2.3pt $k$}}}
\def\qslash{\not{\hbox{\kern-2.3pt $q$}}}

\def\p1{{\bf p}_1}
\def\p2{{\bf p}_2}
\def\k1{{\bf k}_1}
\def\k2{{\bf k}_2}

\newcommand{\dddM}{\kern 0.2em \raise 1.9ex\hbox{$...$}\kern -1.0em \hbox{$M$}}
\newcommand{\dddQ}{\kern 0.2em \raise 1.9ex\hbox{$...$}\kern -1.0em \hbox{$Q$}}
\newcommand{\dddI}{\kern 0.2em \raise 1.9ex\hbox{$...$}\kern -1.0em\hbox{$I$}}
\newcommand{\dddJ}{\kern 0.2em \raise 1.9ex\hbox{$...$}\kern-1.0em
\hbox{$J$}}
\newcommand{\dddcalJ}{\kern 0.2em \raise 1.9ex\hbox{$...$}\kern-1.0em
\hbox{${\cal J}$}}

\newcommand{\dddO}{\kern 0.2em \raise 1.9ex\hbox{$...$}\kern -1.0em
\hbox{${\cal O}$}}
\def\dddz{\raise 1.5ex\hbox{$...$}\kern -0.8em \hbox{$z$}}
\def\dddd{\raise 1.8ex\hbox{$...$}\kern -0.8em \hbox{$d$}}
\def\dddbd{\raise 1.8ex\hbox{$...$}\kern -0.8em \hbox{${\bf d}$}}
\def\ddbd{\raise 1.8ex\hbox{$..$}\kern -0.8em \hbox{${\bf d}$}}
\def\dddx{\raise 1.6ex\hbox{$...$}\kern -0.8em \hbox{$x$}}

\begin{document}

\vspace*{2cm}

\title{Inflationary perturbations  in bimetric gravity}
\author{Giulia Cusin, Ruth Durrer, Pietro Guarato and  Mariele Motta}
\affiliation{D\'epartement de Physique Th\'eorique and Center for Astroparticle Physics, Universit\'e de Gen\`eve, 24 quai Ansermet, CH--1211 Gen\`eve 4, Switzerland}
\email{giulia.cusin@unige.ch, ruth.durrer@unige.ch, pietro.guarato@unige.ch, mariele.motta@unige.ch}

\date{\today}

\begin{abstract}

In this paper we study the generation of primordial  perturbations in a cosmological setting of bigravity  during inflation. We consider a model of bigravity which can reproduce the $\La$CDM background and large scale structure and a simple model of inflation with a single scalar field and a quadratic potential. Reheating is implemented with a toy-model in which the energy density of the inflaton is entirely dissipated into radiation. 
We present analytic and numerical results  for the evolution of primordial perturbations in this cosmological setting. We find that  the amplitude of tensor perturbations generated during inflation is sufficiently suppressed to avoid the effects of the tensor instability discovered in Refs.~\cite{Lagos:2014lca,Cusin:2014psa} which develops during the cosmological evolution in the physical sector. We argue that from a pure analysis of the tensor perturbations this bigravity model is compatible with present observations. However, we derive rather stringent limits on inflation from the vector and scalar sectors.

\end{abstract}


\maketitle

\section{Introduction}
\label{s:intro}

The question whether the graviton may have a mass has attracted considerable attention in the last decade. 
However, constructing a viable theory of massive gravity is a non-trivial problem since the presence of a mass term in the gravity action removes diffeomorphism invariance: the metric has six degrees of freedom (four being absorbed by the Bianchi identities),  five of these represent the massive graviton while the sixth is usually a ghost, the so-called Boulware-Deser ghost~\cite{Boulware:1973my,Deser:1967zzb}. To remove this ghost one needs an additional constraint.  This has been achieved with a very specific form of the potential for the gravitational field, the dRGT (de Rham, Gabadadze, Tolley) potential~\cite{deRham:2010ik,deRham:2010kj,Hassan:2011hr}, which has been the basis for much  of the recent work on this topic (see, e.g.,~\cite{Hassan:2011vm,Hassan:2011tf,Koyama:2011yg,Guarato:2013gba} and refs. therein). 

In massive gravity theories, the mass term is defined with respect to a fixed reference metric and the possible solutions of course strongly depend on this reference metric. Moreover, even when choosing the reference metric to be Friedmann, the resulting solutions either do not show the well known cosmological behavior, or they are unstable~\cite{Gumrukcuoglu:2011zh, D'Amico:2011jj, Langlois:2012hk, Fasiello:2012rw, Solomon:2014iwa}, see~\cite{deRham:2014zqa} for a review.

Also from a theoretical point of view, it is rather unsatisfactory to introduce the reference metric as an  `absolute element', i.e., a non-dynamical field in the theory. For this reason, bimetric (or more general multi-metric) theories, with a dynamical  reference metric, are more natural~\cite{Hassan:2011zd, Hassan:2011ea, Hassan:2012wr}. Investigations of theoretical aspects of bimetric massive gravity can be found in~\cite{Akrami:2014lja, Hassan:2014vja, deRham:2014fha, Cusin:2014zoa, Noller:2014sta, Akrami:2013ffa, Comelli:2013tja, deRham:2014gla}. 

Cosmological solutions of bimetric theories can actually fit the expansion history of the accelerating Universe~\cite{Volkov:2011an, Comelli:2011zm, Konnig:2014dna, Tamanini:2013xia,Fasiello:2013woa}. Observational tests of several  models of bigravity are presented in \cite{Solomon:2014dua,vonStrauss:2011mq,Berg:2012kn,2013JHEP...03..099A,Konnig:2013gxa}. The cosmology of bigravity in various cosmological settings is studied in \cite{Akrami:2015qga,Konnig:2015lfa} while in Refs.~\cite{Gumrukcuoglu:2015nua, Comelli:2015pua, Enander:2014xga}  the cosmology of models of bigravity where matter is coupled to a combination of the two metrics is investigated.

Cosmological perturbations in bigravity have been studied in different settings and for different models in \cite{Comelli:2012db, Khosravi:2012rk, Comelli:2014bqa, DeFelice:2014nja}. A more generic study of instabilities in bimetric theories can be found in Refs.~\cite{Kuhnel:2012gh,Lagos:2014lca}. Recently, scalar perturbations of these models have been investigated and it has been shown that there exists a class of models of bigravity that admit solutions with scalar perturbations free of exponential instabilities at all times, while the other models do exhibit exponential instabilities in the scalar sector~\cite{Amendola_pert, Konnig:2015lfa}.

The evolution of tensor perturbations in this particular class of models has been first studied in \cite{Lagos:2014lca} and in more detail in \cite{Cusin:2014psa}. The problem of how cosmological observations  are affected by these tensor instabilities and possible ways out  are discussed in~\cite{Amendola:2015tua}. In \cite{Fasiello:2015csa}  a general analysis of the tensor sector in models which are free from known instabilities is presented and it is discussed how measurements of the amplitude of primordial gravitational waves can be used to constrain them.

In Ref.~\cite{Cusin:2014psa} it has been found that tensor perturbations are affected by a power-law instability connected with the violation of the Higuchi bound~\cite{Higuchi:1986py} in the sector not-coupled to matter. This instability is then transferred to the physical sector through the coupling between the two tensor modes. By fine-tuning the amplitude of the unstable tensor mode to be highly suppressed with respect to the one of the physical sector at the end of inflation, one can achieve that the instability does not show up in the physical metric until today. In \cite{Cusin:2014psa} this fine-tuning is explicitly quantified. 
The problem of the viability of the model is therefore translated into the question: is the  amplitude of the uncoupled tensor mode after inflation sufficiently suppressed with respect to the one of the graviton coupled to matter?

In this paper we address this question in detail, i.e., we embed the model of bigravity studied in  \cite{Cusin:2014psa} in an inflationary scenario and determine the amplitude and the spectrum of primordial tensor perturbations generated at the end of inflation. We find an expression for the ratio between the amplitude of the two tensor modes at the beginning of the radiation era as a function of the reheating temperature of the inflationary model. We argue that however, the amplitude of the non-physical metric is nearly entirely in the constant mode and the growing mode amplitude is much smaller, actually too small to affect the  `physical metric', i.e. the metric which is coupled to matter in the way discussed in  \cite{Cusin:2014psa}.  From this linear analysis, the model is therefore not in conflict with observations. 

We also investigate the vector and the scalar sector. We find that in the model considered, in addition to the nearly scale invariant inflaton mode, very large vector perturbations and a very red spectrum of scalar perturbations are generated during inflation. The condition for the vector mode not to spoil perturbation theory (i.e. back-react) during inflation constrains the scale of inflation substantially.

While we were working on this, a  study of primordial gravitational waves in this model has appeared in~\cite{Johnson:2015tfa} in a larger context. Our analysis goes beyond the results presented in~\cite{Johnson:2015tfa}. We perform an analytical study of primordial perturbations in all the sectors. We study in detail, both numerically and analytically, tensor perturbations during inflation and reheating. The main results of~\cite{Johnson:2015tfa} are confirmed. We also find that tensor perturbations of the second metric which are generated during inflation are too small to affect the observable gravitational waves in the physical metric. 

The  paper is organised as follows. In the next section we present the equations of motion of bimetric gravity for cosmological (i.e. homogeneous and isotropic) spacetimes. We then specialise to a model which gives an acceptable expansion history. In  Sec.~\ref{inflation} we present our model of inflation and reheating and we study its background evolution.  In Sec.~\ref{s:pert} we briefly review the perturbation equations of bimetric gravity. The study of tensor perturbations is presented in Sec.~\ref{tensors} and  discussed in Sec.~\ref{discussion} . In Sec.~\ref{s:vector} we study the generation of vector perturbations during inflation and in Sec.~\ref{s:scalar} we discuss scalar perturbations. Finally, in Sec.~\ref{conclusion} we  conclude.\\

{\bf Notation:} We set $c=\hbar=k_{\rm Boltzmann}=1$. $M_g=1/\sqrt{8\pi G}\equiv M_p\simeq 2.4\times 10^{18}$GeV is the reduced Planck mass. 
We work with the metric signature $(-, +, +, +)$, and we restrict to $4$ spacetime dimensions. With  $\cdot $ and  with $'$  we indicate derivatives with respect to physical time and to conformal time, respectively. The conventions for the bigravity action are those of  ~\cite{Hassan:2011zd, Hassan:2011ea, Hassan:2012wr}. We consider only one of the two metrics coupled to matter,  and we restrict to minimal couplings.  For self-consistency, the  Hassan-Rosen bi-metric action and the general equations of motion are given explicitly  in Appendix \ref{general setting}.

\section{Cosmological ansatz and  background equations}

We consider solutions of bigravity where both metrics are spatially isotropic and homogeneous. For simplicity, we also assume that both metrics have  flat spatial sections, $K=0$. Modulo time reparameterizations,  the most general form for the metrics (in conformal time $\tau$) is
\be
g_{\mu\nu}dx^{\mu}dx^{\nu}=a^2(\tau)\left(-d\tau^2+\delta_{ij}dx^idx^j\right)\,,
\ee
\be
f_{\mu\nu}dx^{\mu}dx^{\nu}=b^2(\tau)\left(-c^2(\tau) d\tau^2+\delta_{ij}dx^idx^j\right)\,.
\ee
Here $a$ and $b$ are the scale factors of the two metrics and $c$ is a lapse function for $f$.
It is convenient to define both the conformal Hubble parameter ($\HH$) and the standard one (${H}$) for both metrics
\be
H=\frac{\mathcal{H}}{a}=\frac{a'}{a^2}\,,\hspace{0.5 cm} H_f=\frac{\mathcal{H}_f}{b}=\frac{b'}{b^2\,c}\,,
\ee
where $'$  denotes the derivative with respect to the conformal time $\tau$. We introduce also the ratio between the two scale factors
\be
r=\frac{b}{a}\,.
\ee

In the matter sector, we consider the energy-momentum tensor of a covariantely conserved perfect fluid with equation of state $p=w \rho$ and 4-velocity $u^{\mu}$. Explicitly,
\begin{align}
&T_{\mu\nu}=\left(p+\rho\right)\, u_{\mu} u_{\nu}+p \,g_{\mu\nu}\,,\\
&\rho'=-3\,\mathcal{H}\,(\rho+p),\\
&p=w\rho\,.
\end{align}
The general Lagrangian of bimetric gravity and the resulting modified Einstein equations for the metrics $g$ and $f$ are presented in the Appendix~\ref{general setting}.
 
The Bianchi constraint in the cosmological ansatz can be written as
\be\label{e:bian}
\rho_G'=-3\,\mathcal{H}\,\left(\rho_G+p_G\right)\,,
\ee
where we have introduced a `gravity fluid' with density and pressure given by
\be
\rho_G=\frac{m^2}{8\pi G}\left(\beta_3\, r^3+3\beta_2\,r^2+3\beta_1\,r+\beta_0\right)\,,
\ee
\be
p_G=-\frac{m^2}{8\pi G}\left(\beta_3 c \,r^3+\beta_2(2c+1)r^2+\beta_1(c+2)r+\beta_0\right)\,.
\ee
Here the $\beta_i$ are the parameters of the bigravity potential, see Appendix~\ref{general setting}.
It is easy to show that the Bianchi constraint (\ref{e:bian}) is equivalent to
\be \label{Bianchic}
m^2\left(\beta_3\,r^2+2\beta_2\,r+\beta_1\right)\,(c\,b\,a'-a\,b')=0\,.
\ee
The equations of motion (the Friedmann equation and the acceleration equation) for the metric $g$  are
\be\label{F11}
3H^2=8\pi G\,\left(\rho+\rho_G\right)\,,
\ee
\be\label{F12}
3H^2+\frac{2H'}{a}=-8\pi G\,\left(p+p_G\right)\,,
\ee
while for the $f$ metric we find the equations of motion
\be\label{F21}
3\,H_f^2=\frac{m^2}{\al^2}\left(\frac{\beta_1}{r^3}+\frac{3\beta_2}{r^2}+\frac{3\beta_3}{r}+\beta_4\right)\,,
\ee
\be\label{F22}
H_f'=\frac{m^2}{\al^2}\left(\frac{1}{c}-1\right) \left(\frac{\beta_1}{r^3}+\frac{2\beta_2}{r^2}+\frac{\beta_3}{r}\right)\, ,
\ee
where $\al$ is a dimensionless parameter, $\al\equiv M_f/M_g$ (see also Appendix \ref{general setting}).
Under the rescaling $f_{\mu\nu}\rightarrow \al^{-2}\,f_{\mu\nu}$ and $\beta_n\rightarrow \al^n\,\beta_n$, the equations of motion become independent of $\al$ \cite{Berg:2012kn,Hassan:2012wr}, which has motivated many works on the cosmology of bigravity to simply set $\al=1$, as we shall do here. Recently, however, it has been argued that this choice actually hides the possibility to recover General Relativity (GR) with a cosmological constant in the limit $\al\ra 0$, see \cite{Akrami:2015qga} for a detailed discussion.

We distinguish two branches of solutions, depending on how the Bianchi constraint (\ref{Bianchic}) is implemented. Either there is an algebraic constraint for $r$,
\be
\beta_3\,r^2+2\beta_2\,r+\beta_1=0\,,
\ee
or 
\be\label{Bianchiconstraint1}
\mathcal{H}_f=\mathcal{H}\, , \qquad rH_f = H \,.
\ee
At the background level the first branch with constant $r$ is equivalent to  GR with an effective cosmological constant, while the second one gives rise to a richer cosmology. We will focus on the second branch in the rest of this work. The Bianchi constraint in the second branch can be re-written as
\be\label{Bianchiconstraint}
c=\frac{r'+r\HH}{r\HH}\,.
\ee 
This fixes $c$ as a function of $\mathcal{H}$, $r$ and $r'$.

From now on, we will focus on the so-called `$\beta_1 \beta_4$ model' of bigravity, where all the $\beta_n$ parameters but $\beta_1$ and $\beta_4$ are set to zero. This model is also called the `infinite branch $\beta_1 \beta_4$ model' or `infinite branch bigravity' in Ref. \cite{Amendola_pert}, referring to the fact that the initial condition for $r$ has to be chosen in such a way that $r$ evolves from infinity to a finite value during the cosmological evolution, in order for the exponential instabilities in the scalar sector not to show up. As already mentioned, this model is the only one free of these instabilities. The study of the cosmological evolution of this model has been addressed in a series of recent papers \cite{Cusin:2014psa,Lagos:2014lca,Amendola_pert}.

\section{Scalar field inflation and reheating}\label{inflation}
\subsection{General setting}
In this work we focus on the evolution of the $\beta_1 \beta_4$ model of bigravity during the inflationary period, where the dynamics of the universe is dominated by a scalar field $\phi$, the inflaton, minimally coupled to the physical metric $g$. We consider a simple model of inflation with a  single scalar field with mass $M_{\phi}$ and quadratic potential. We choose the best-fit values $\beta_1m^2=0.48\,H_0^2$ and $\beta_4m^2=0.94\,H_0^2$ obtained in \cite{Konnig:2013gxa} and \cite{Amendola_pert}  by fitting measured growth data and type Ia supernovae.  

The Lagrangian density for the inflaton can be written as
\be
\mathcal{L}_{\phi}=-\frac{1}{2}\partial_{\mu}\phi\,\partial^{\mu}\phi-V(\phi)\,,\hspace{1.4 cm} V(\phi)=\frac{1}{2}M_{\phi}^2\,\phi^2\,.
\ee
The field $\phi$ can in principle interact with other fields such as fermions, gauge bosons, etc., but we assume that this interaction can be neglected during inflation and that energy and pressure are dominated by the contribution from the inflaton. The energy-momentum tensor of $\phi$ is given by
\begin{align}
T_{\mu\nu}=&\,\partial_{\mu} \phi\,\partial_{\nu}\phi+g_{\mu\nu}\,\mathcal{L}_{\phi}=\partial_{\mu}\phi\,\partial_{\nu}\phi-g_{\mu\nu}\left(\frac{1}{2}\,g^{\alpha\beta}\partial_{\alpha}\phi\,\partial_{\beta}\phi+V(\phi)\right)\,.
\end{align}
For the energy density and pressure this yields
\be\label{e:rhophi}
\rho_{\phi}=-T_{0}^{0}=\frac{\phi'^2}{2 a^2}+\frac{1}{2 a^2}\left(\nabla\phi\right)^2+V(\phi)\simeq \frac{\phi'^2}{2 a^2}+V(\phi)\simeq V(\phi) \,,
\ee
and
\be\label{state}
p_{\phi}\equiv\omega_{\phi}\,\rho_{\phi}=\frac{T_{i}^{i}}{3}=\frac{\phi'^2}{2 a^2}-\frac{1}{6 a^2}\left(\nabla\phi\right)^2-V(\phi)\simeq \frac{\phi'^2}{2 a^2}-V(\phi)\simeq -V(\phi) \,.
\ee
The first approximation in eqs.~(\ref{e:rhophi},\ref{state}) is due to the fact that here we suppose that there exists some (sufficiently large) region of space within which we may neglect the spatial derivatives of $\phi$ at some initial time $\tau_i$, explicitly  $\nabla\phi(\bf{x}, \tau_i)\ll \phi'(\bf{x}, \tau_i)$. The second approximation is due to the fact that we also suppose that in this region of space also the time derivative is much smaller than the potential, $\dot{\phi}(\bf{x}, \tau_i)\ll$\, $V^{1/2}(\phi)$. These slow-roll conditions are such that we have $p_{\phi}\equiv\omega_{\phi}\,\rho_{\phi}\simeq -\rho_{\phi}$ and $\rho_{\phi}+3 p_{\phi}\simeq -2\,V(\phi)<0$. 

At early times in some sufficiently large patch, the Universe is dominated by the potential of a slowly varying (slow rolling) scalar field, and hence it is in an inflationary phase. As time goes on, the scalar field starts evolving faster and inflation eventually comes to an end when the time derivative of $\phi$ grows to the order of $V^{1/2}$. When inflation ends, $\phi$ decays rapidly and starts oscillating about its minimum. At the end of inflation, the inflaton oscillates as
\be
\phi\simeq\phi_0\,\cos(M_{\phi}\,\tau)\,.
\ee
The field $\phi$ is a damped harmonic oscillator with frequency $M_{\phi}$. For a harmonic oscillator, when averaging over one period we have
\be
\langle V\rangle =\frac{\langle \phi'^2\rangle}{2 a^2}\,,
\ee
so that 
\be
\langle p_{\phi}\rangle=\left\langle \frac{\phi'^2}{2 a^2}-V\right\rangle=0\,,\hspace{0.4 cm} \text{and hence \quad } \langle \rho_{\phi}\rangle \propto a^{-3}\,.
\ee
We assume that during these oscillations the coupling of $\phi$ to other degrees of freedom than $g_{\mu\nu}$ becomes relevant and the inflaton finally decays into a mix of elementary particles which rapidly thermalize. As a simple approximation to this complicated and model dependent reheating process, we  describe the coupling with the other degrees of freedom by means of a dissipative term $\propto \Gamma \phi '$ in the equation of motion. In physical time, the equation of motion for the inflaton becomes
\be\label{eqm}
\ddot{\phi}+3 H \dot{\phi}+\Gamma \dot{\phi}=-V_{,\phi}(\phi)\,.
\ee
During inflation $H \gg \Gamma$ and particle production is negligible. When $H\simeq \Gamma$, reheating takes place and the inflaton energy is rapidly dissipated into other particles which couple to the inflaton. 

We consider a toy-model of reheating in which the energy density of the inflaton is entirely dissipated into radiation. In this setting, the total energy momentum tensor has a contribution given by the inflaton and one given by radiation,
\be
T_{\mu\nu}=T_{\mu\nu}^{(\phi)}+T_{\mu\nu}^{(r)}\,.
\ee
Initially $T_{\mu\nu}^{(r)}=0$.
The total energy momentum is covariantly conserved:
\be\label{ccons}
\nabla_{\mu} T^{\mu\nu}=0\, \qquad \Longrightarrow \qquad \nabla_{\mu} T^{\mu\nu\,(\phi)}=-\nabla_{\mu} T^{\mu\nu\,(r)}\,.
\ee
In our cosmological setting, eq. (\ref{ccons}) is equivalent to the following set of equations:
\be\label{i}
\dot{\rho_{\phi}}+3\,\frac{\dot{a}}{a}\,\left(\rho_{\phi}+p_{\phi}\right)=-\Gamma\,\left(\rho_{\phi}+p_{\phi}\right)\,,
\ee
\be\label{ii}
\dot{\rho_{r}}+3\,\frac{\dot{a}}{a}\,\left(\rho_{r}+p_{r}\right)=\Gamma\,\left(\rho_{\phi}+p_{\phi}\right)\,.
\ee
It is easy to check that eq. (\ref{i}) is equivalent to the equation of motion for the inflaton, eq. (\ref{eqm}). 

\subsection{Background evolution during inflation}\label{background}
To study the background evolution during inflation, we consider as a complete set of independent equations the two Friedmann equations,  (\ref{F11}), (\ref{F21}), and the acceleration equation for the $g$-metric, (\ref{F12}), the Bianchi constraint in the second branch, (\ref{Bianchiconstraint1}), the equation of state for the inflaton,  (\ref{state}) and the equation of motion for the inflaton, (\ref{eqm}).  Solving the two Friedmann equations together with the acceleration equation for $g$, we can express $r'$, $H$ and $\rho_{r}$ as functions of $r$ and $\rho_\phi$
\be\label{rdot}
\frac{\dot r}{H}=\frac{r'}{\mathcal{H}}=\frac{-3\,r\left(1+\omega_{r}\right)\left(\beta_4\,r^3-3\beta_1 r^2+\beta_1\right)+3r^2\left(\omega_{r}-\omega_{\phi}\right) 8\pi G\,m^{-2}\,\rho_{\phi}}{2\beta_4\,r^3-3\beta_1 r^2-\beta_1}\,,
\ee
\be\label{H}
H^2=\frac{\HH^2}{a^2}=m^2\,\frac{\beta_1+\beta_4 r^3}{3\,r}\,,
\ee
\be\label{rhoreq}
8 \pi G \rho_r=m^2\frac{\beta_1}{r}-3\,m^2 \beta_1 r+m^2 \beta_4\,r^2-8\pi G \rho_{\phi}\,,
\ee
where the suffixes  `$r$' and  `$\phi$' refer to radiation and to the inflaton, respectively;  with $\rho_{\phi}=\frac{1}{2 a^2}\,\phi'^2+\frac{1}{2}M_{\phi}^2\phi^2$ and $\omega_{\phi}\equiv p_{\phi}/\rho_{\phi}=-1+\phi'^2/(a^2 \rho_{\phi})$. The equation of motion for the inflaton, eq. (\ref{eqm}), in conformal time can be written as
\be\label{phieq}
\phi''+2\mathcal{H}\phi'+a \Gamma \phi'+a^2 V_{,\phi}(\phi)=0\,.
\ee

The two differential equations (\ref{rdot}) and (\ref{phieq}) are coupled and we solve them together with initial conditions given deep in the inflationary epoch. We choose the expectation value of the inflaton at the beginning of inflation to be of order $10\,M_p$.  Since during inflation the slow-roll condition holds and $\Gamma \ll H$, the initial conditions for eq. (\ref{phieq}) can then be parametrized as
\be\label{inphi}
\phi(\tau_{i})=10 M_p\,,
\ee
\be\label{inphiprime}
\frac{\phi'}{a} (\tau_{i})=-\frac{V_{,\phi}}{3 H}(\tau_{i})=-\frac{M_{\phi}^2\, \phi}{3 H}(\tau_{i})\,,
\ee
where we choose for the mass of the inflaton $M_{\phi}\simeq 0.2$ eV.\footnote{Therefore, from $H(\tau_{i})^2\simeq \frac{8\pi G}{3}\,V(\tau_{i})\simeq \left(\frac{M_{\phi}}{M_p}\right)^2\,\frac{\phi(\tau_{i})^2}{6}$ it follows that 
$H(\tau_{i})\simeq 1$ eV.} Therefore, the state parameter for the inflaton can also be written as
\be\label{21}
\omega_{\phi}|_{\tau\approx\tau_{i}}=-1+\frac{2M_{\phi}^2}{3\beta_4m^2r^2}\, ,
\ee
where in the last equality we have used the fact that during inflation $r\gg 1$ (as follows from eq. (\ref{H})) and $\rho_r\simeq0$. 

Eq. (\ref{H}) and $r\gg 1$  also imply that during inflation
\be\label{rin}
r(\tau_i)\simeq \sqrt{\frac{3\,H(\tau_{i})^{2}}{m^{2}\,\beta_4}}~\sim ~ {\cal O}\left(\frac{H}{H_0}\right)\,.
\ee
Once the coupled differential equations  (\ref{rdot}) and (\ref{phieq}) are solved with initial conditions (\ref{inphi}), (\ref{inphiprime}) and (\ref{rin}),  the evolution of the Hubble parameter, of the lapse $c$ and of $\rho_r$  (via eq. (\ref{rhoreq})) can be derived. 

\begin{figure}[ht!]
    \centering
     \subfigure[\label{r}]
     {\includegraphics[scale=0.37]{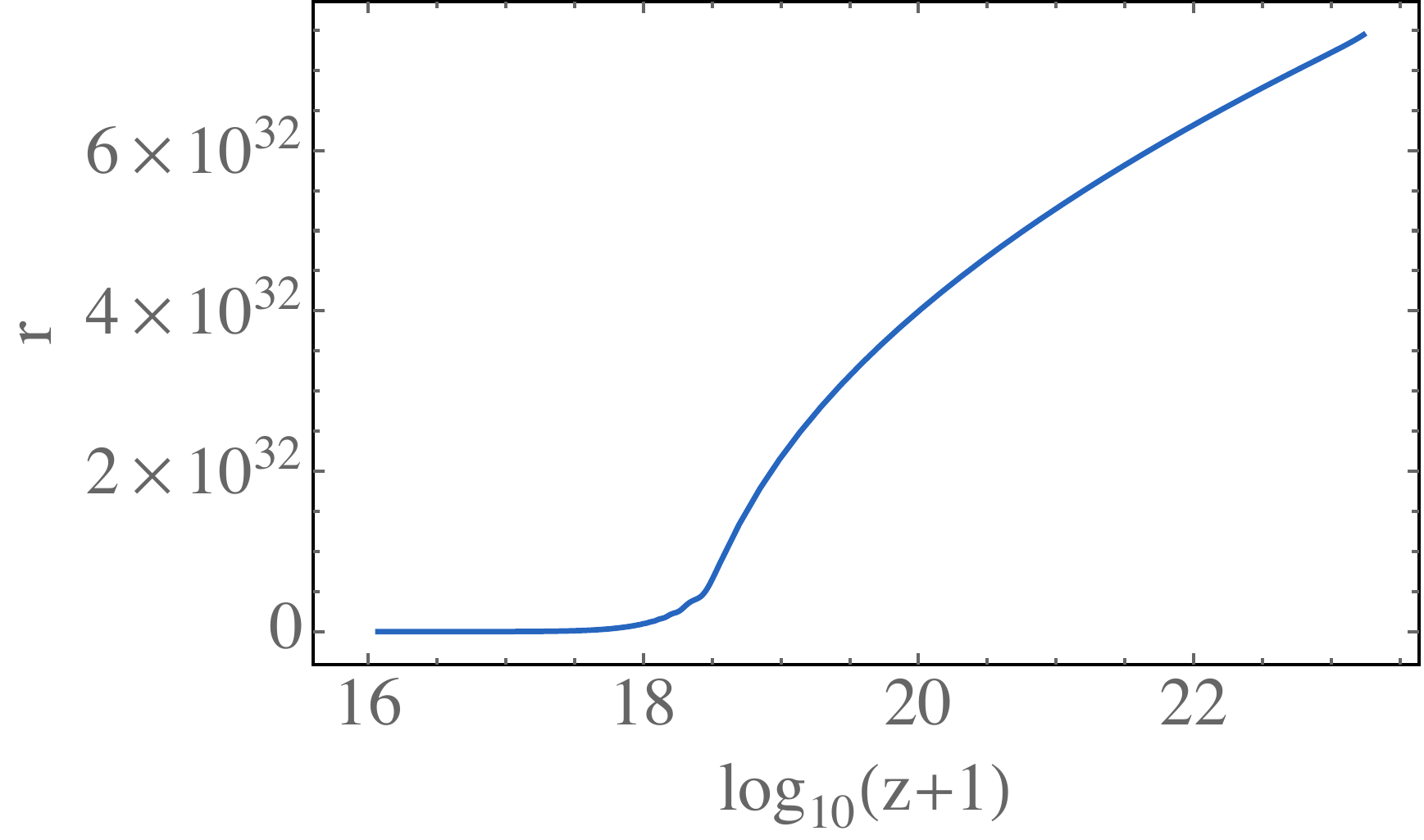}}\qquad\qquad\qquad
 \subfigure[\label{logr}]
     {\includegraphics[scale=0.37]{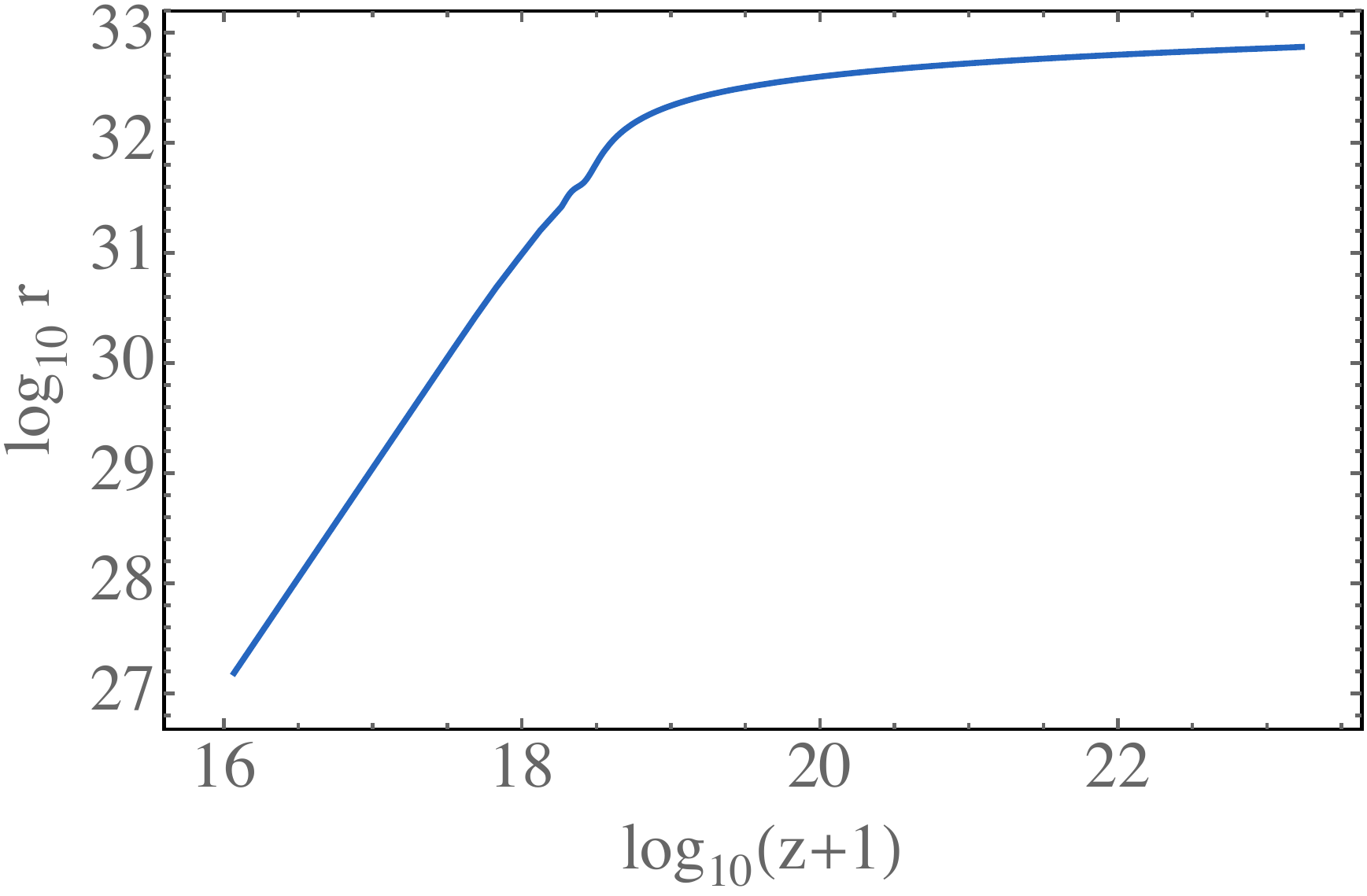}}
         \subfigure[\label{c}]
     {\includegraphics[scale=0.37]{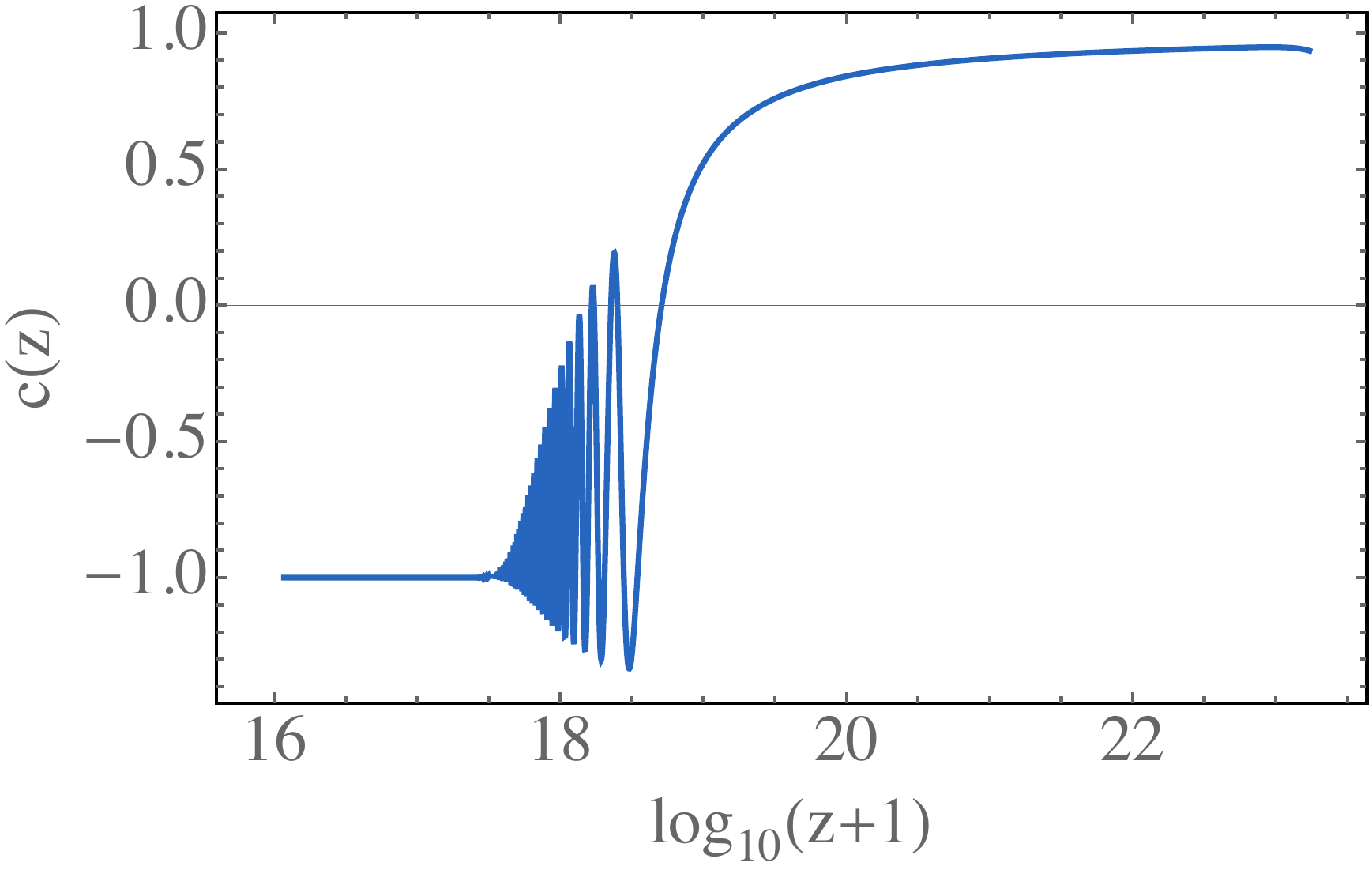}}\qquad\qquad\qquad
              \subfigure[\label{omega}]
     {\includegraphics[scale=0.37]{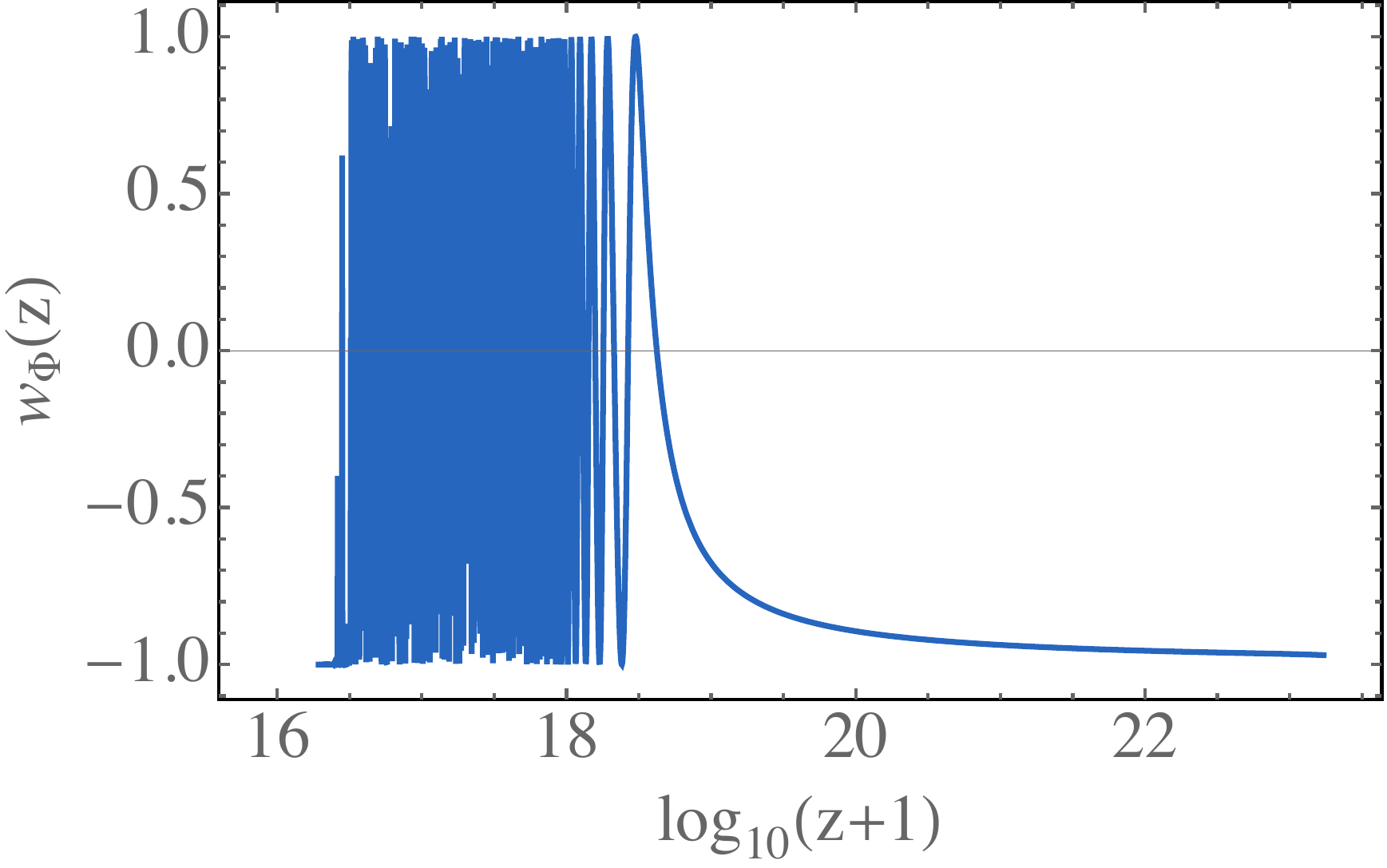}}
 \subfigure[\label{phiz}]
     {\includegraphics[scale=0.37]{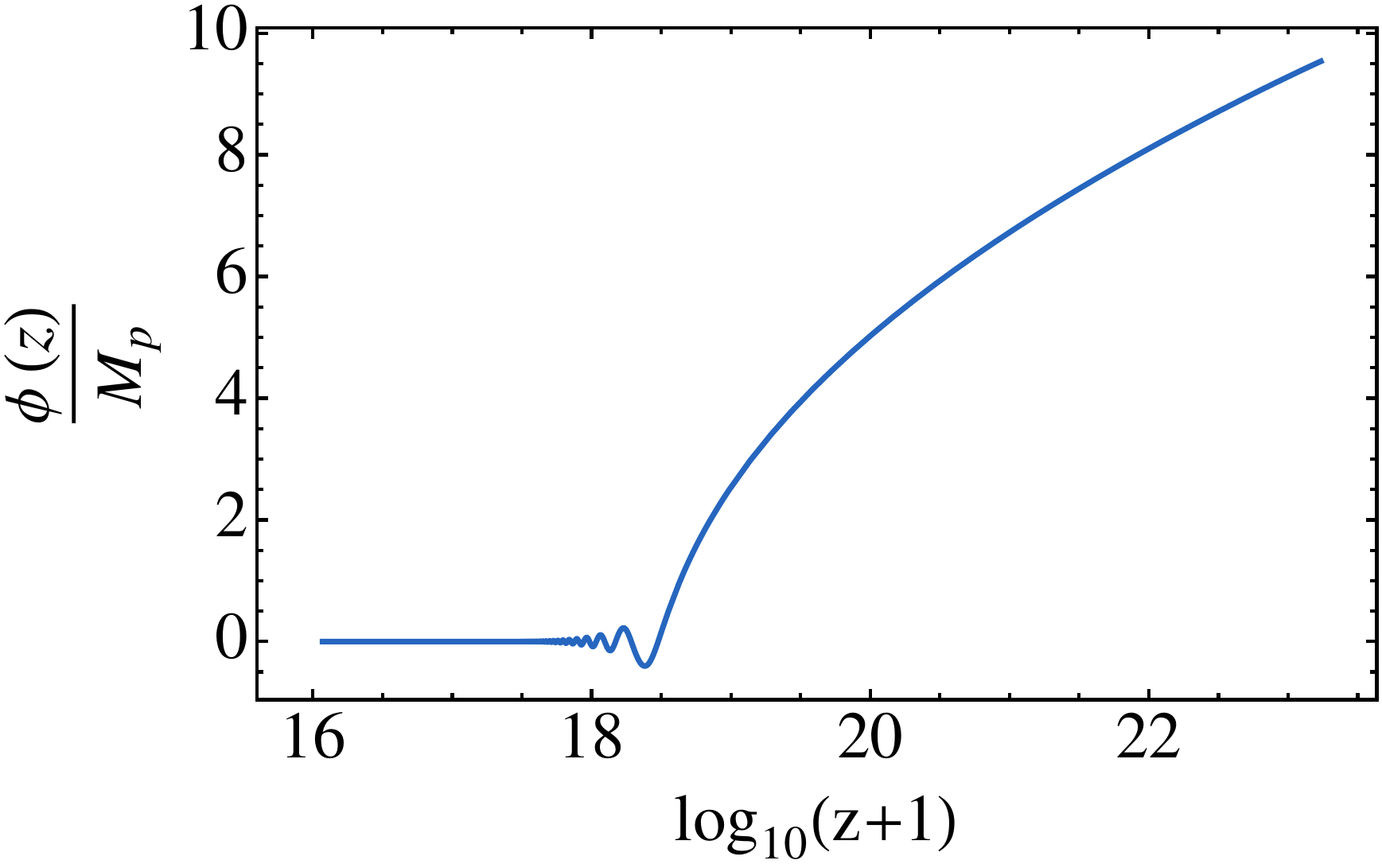}}\qquad\qquad\qquad
 \subfigure[\label{lohphiz}]
      {\includegraphics[scale=0.37]{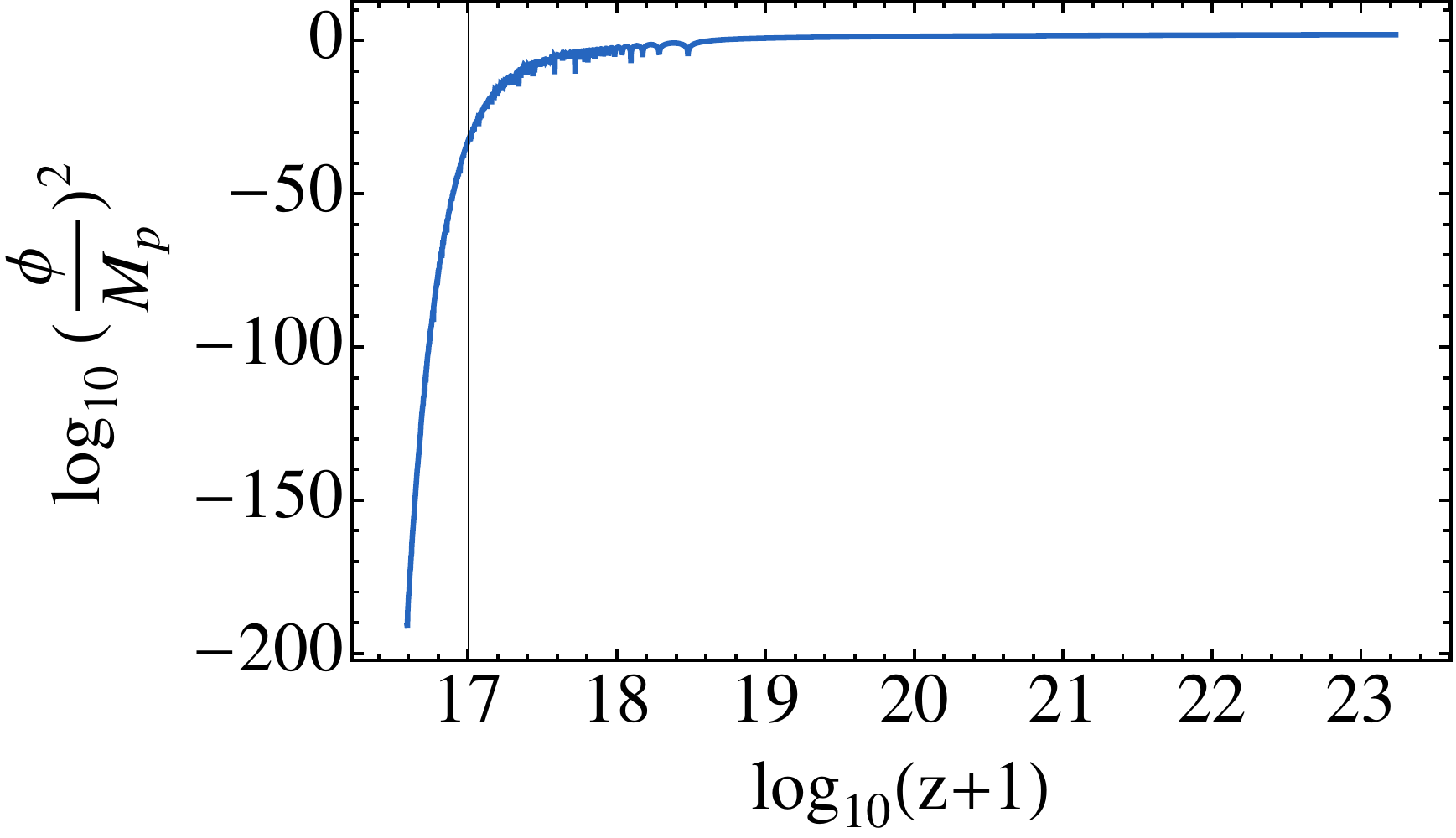}}
 \caption{\label{fig1}We show the ratio between the scale factors $r$, the lapse of the $f$-metric $c$, the equation of state parameter of the inflaton $\omega_{\phi}$  and  the expectation value of the inflaton as functions of redshift during inflation. We have chosen 
 $\phi(z_{i})=10\,M_p$ and $H(z_{i})=1\,$ eV. 
 The parameter  $\Gamma$ in eq. (\ref{phieq}) has been chosen  such  that $\Gamma=H(z_{\rm reh})$ with $z_{\rm reh}=5\cdot 10^{17}$. Note how the oscillations of $\phi$ lead to strong oscillations of the lapse function $c$ and the equation of state parameter $\om_\phi$ at the end of inflation.}
 \end{figure}

\begin{figure}[ht!]
    \centering
     \subfigure[\label{HHH}]
     {\includegraphics[scale=0.37]{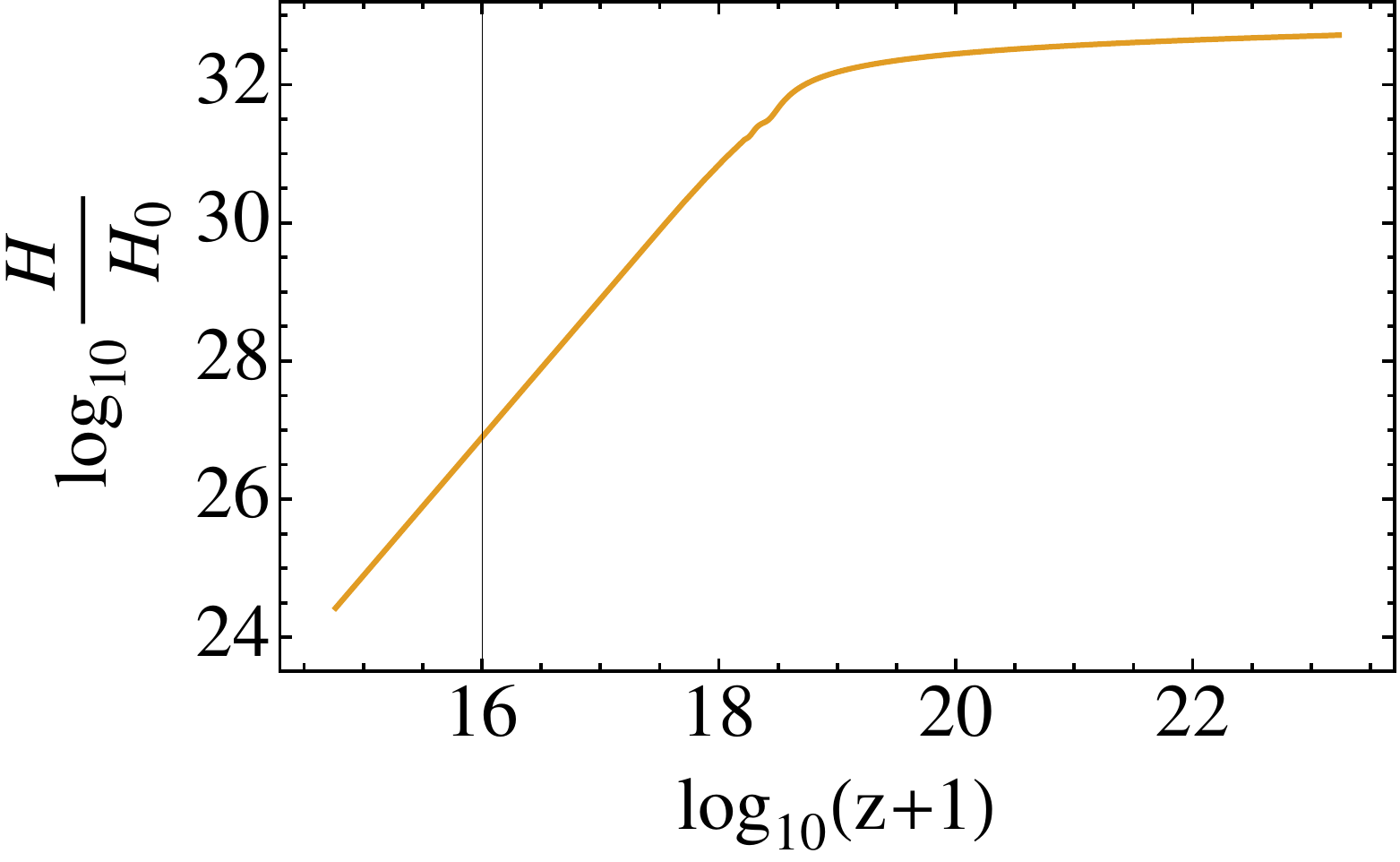}}\qquad\qquad\qquad
 \subfigure[\label{HH}]
     {\includegraphics[scale=0.37]{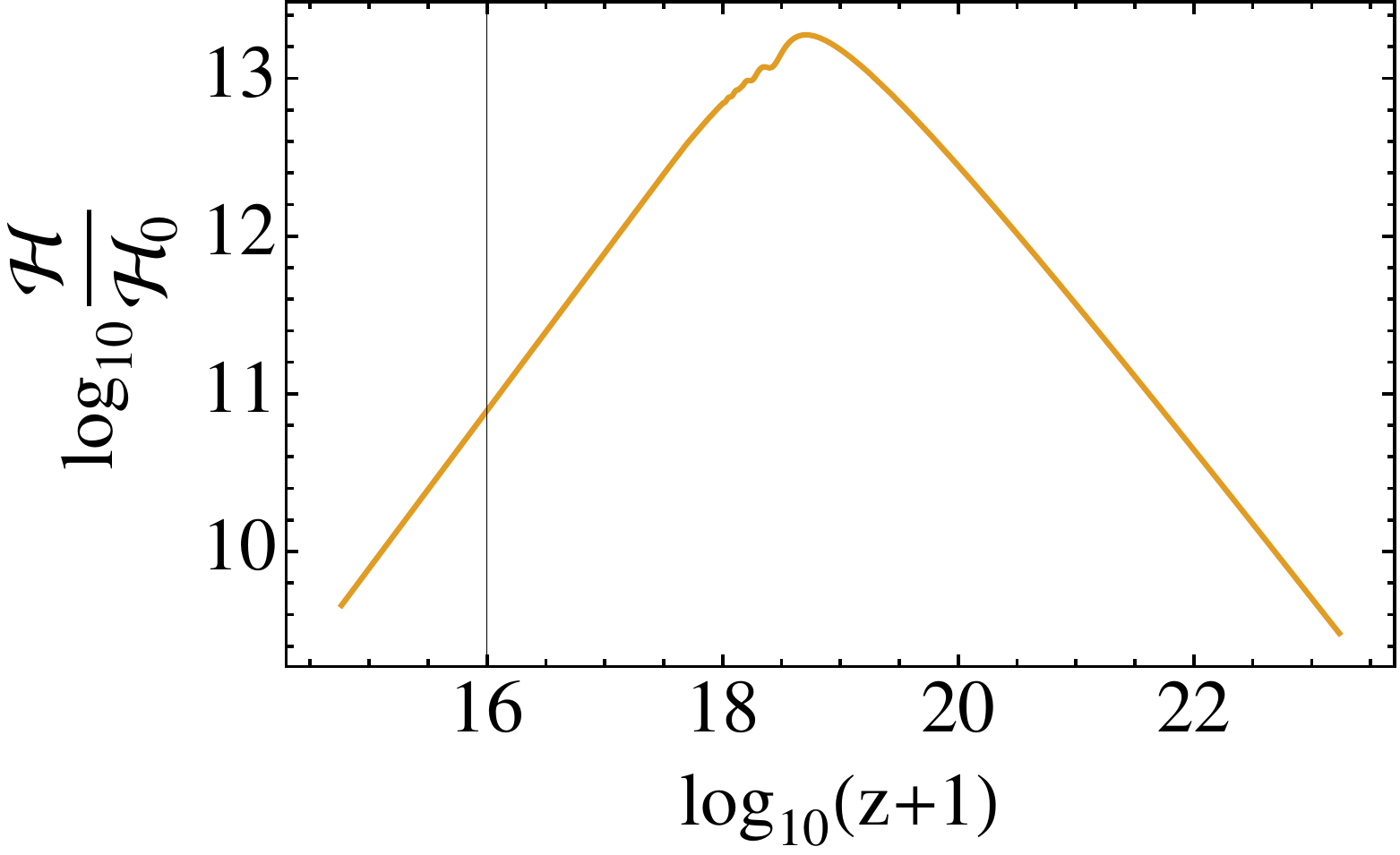}}
 \caption{\label{fig2} The physical and the conformal Hubble parameters as functions of redshift, panels \ref{HHH} and \ref{HH} respectively. }
 \end{figure}

\begin{figure}[ht!]
    \centering
     \subfigure[\label{rhophi}]
     {\includegraphics[scale=0.36]{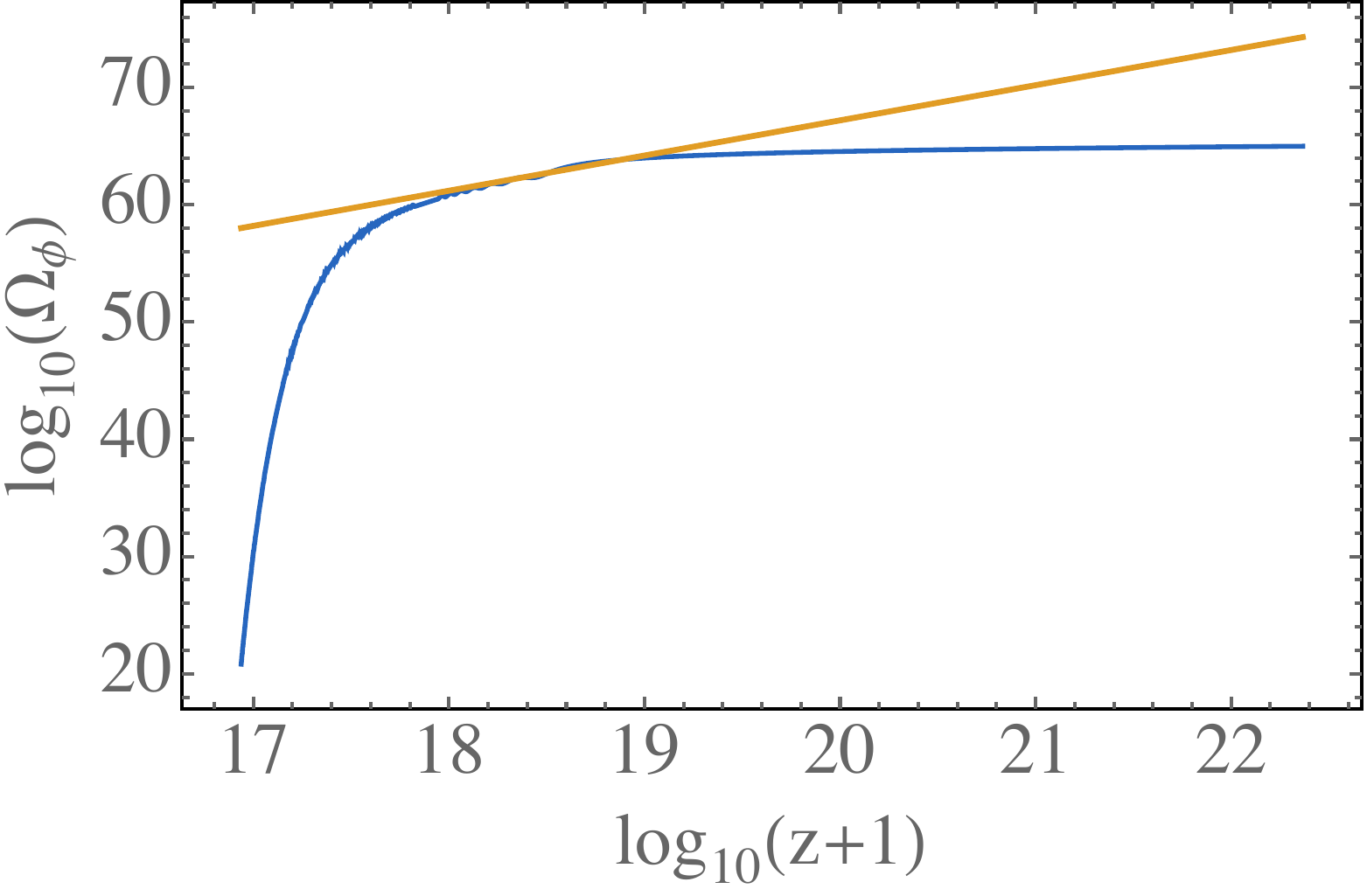}}\qquad\qquad\qquad
 \subfigure[\label{rhor}]
     {\includegraphics[scale=0.36]{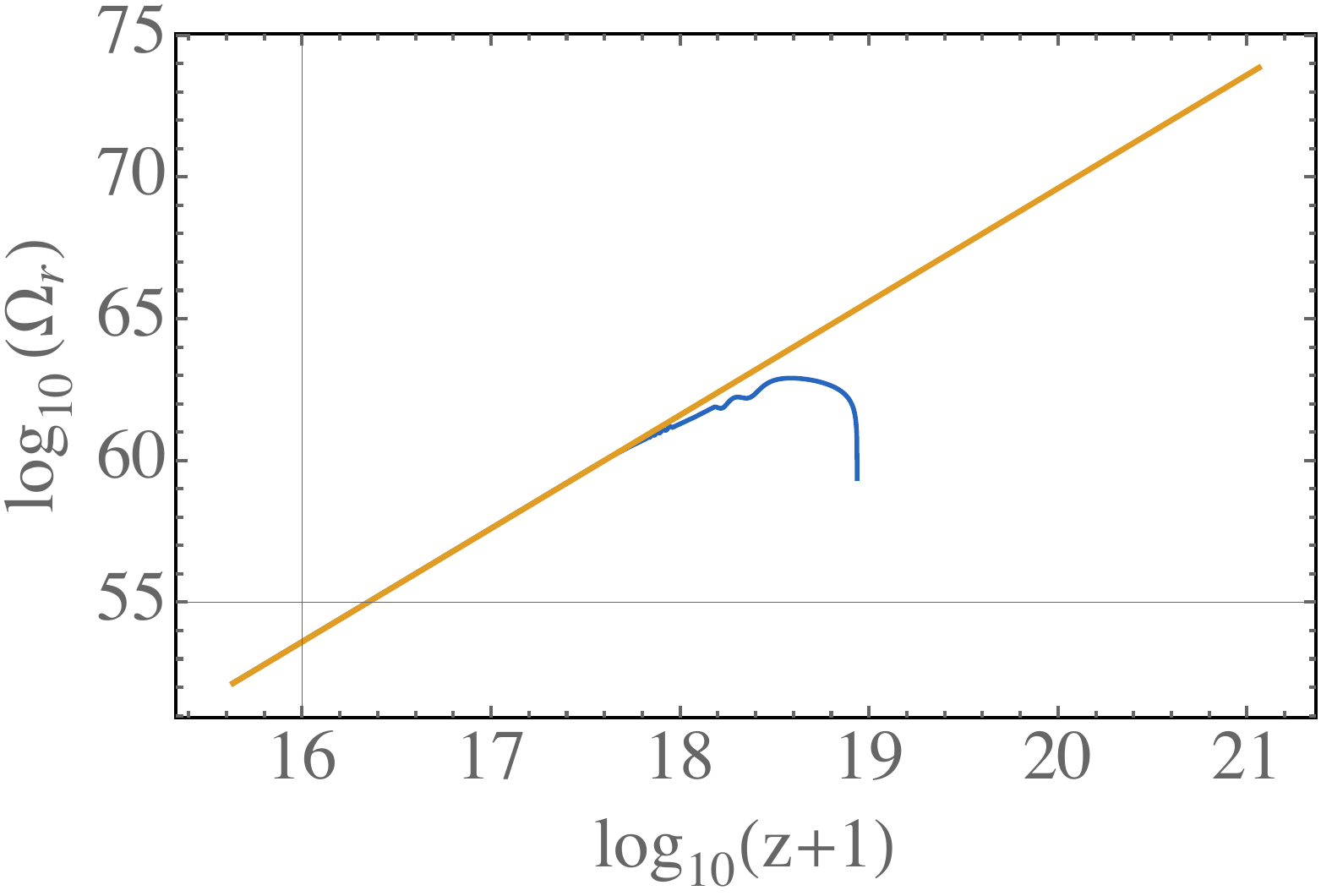}}
 \caption{\label{fig3} Evolution of the energy density of the inflaton and of radiation (in blue), normalized with respect to the critical energy density of the universe today. The yellow curves in panels~\ref{rhophi} and \ref{rhor} are $\propto a^{-3}$ and $\propto a^{-4}$, respectively. }
 \end{figure}

The results of the numerical integration are shown in Figs. \ref{fig1}, \ref{fig2} and \ref{fig3}. For the numerical integration, the parameter  $\Gamma$ in eq. (\ref{phieq}) has been chosen  such  that $\Gamma=H(z_{\rm reh})$, where $z_{\rm reh}=5\cdot 10^{17}$ is the reheating redshift. Fig. \ref{fig1} shows that the inflaton starts oscillating at the end of inflation, and that this oscillation is transferred to $\omega_{\phi}$ and $c$, which starts from the value $c=1$ during inflation and becomes $c=-1$ in radiation domination. The variable $r$ is almost constant during inflation ($r_{I}\sim 10^{33}$) and  it starts to decay rapidly in the radiation dominated era. Fig. \ref{HHH} shows that the physical Hubble parameter is almost constant during inflation and then starts to decrease. Fig. \ref{fig3} shows that at the end of inflation the energy density of the inflaton is matter-like while the energy density of radiation produced by the decaying of the inflaton has the usual evolution with time\footnote{We have also checked that the evolution of $\rho_r$ from eq. (\ref{rhoreq}) is equivalent to the one obtained solving the differential equation (\ref{ii}), with vanishing initial condition for $\rho_r$ at early times.} $\propto a^{-4}$.

\section{Analysis of perturbations: gauge invariant variables}\label{s:pert}

We consider perturbations around the Friedmann backgrounds,
\be
g_{\mu\nu}=\bar{g}_{\mu\nu}+a^2\, h_{g\,\mu\nu}\,,\hspace{1.5cm} f_{\mu\nu}=\bar{f}_{\mu\nu}+b^2\, h_{f\,\mu\nu}\,.
\ee
In this section, background quantities are indicated with  an overbar. We parametrize the perturbations in as follows:
\be
\left(h_{g\,\mu\nu}\right)=\left(
\begin{array}{cc}
-2 A_g&C_{g j}-\partial_j B_g\\
C_{g i}-\partial_i B_g& h_{g ij}^{TT}+\partial_i\mathcal{V}_{gj}+\partial_j\mathcal{V}_{gi}+2\partial_i\partial_j E_g+2\delta_{ij} F_g\\
\end{array}
\right)\,,
\ee
 \be
\left(h_{f\,\mu\nu}\right)=\left(
\begin{array}{cc}
-2 c^2 A_f&C_{f j}-\partial_j B_f\\
C_{f i}-\partial_i B_f& h_{f ij}^{TT}+\partial_i\mathcal{V}_{fj}+\partial_j\mathcal{V}_{fi}+2\partial_i\partial_j E_f+2\delta_{ij} F_f\\
\end{array}
\right)\,,
\ee
with
\be
\partial_i C_{g,f}^i=\partial_i\mathcal{V}_{g,f}^i=\partial_i h_{g,f}^{TT ij}=0\,,\hspace{1cm} \delta^{ij}h_{g,f ij}^{TT}=0\,.
\ee
Spatial indices are raised and lowered using the flat  spatial metric $\de_{ij}$.

In the scalar sector we have 8 fields and 2 gauge freedoms, hence we can form 6 gauge invariant combinations which can be chosen as~\cite{Comelli:2012db,Solomon:2014dua} 
\bees
&\Psi_g=A_g-\mathcal{H}\,\Gamma_g A_g-\Gamma_g'\,,\hspace{0.5 cm}\Psi_f=A_f+c^{-2}\left(\frac{c'}{c}-c\,\mathcal{H}_f\right)\,\Gamma_f-c^{-2}\,\Gamma_f'\,,\nn\\
&\Phi_g=F_g-\mathcal{H}\,\Gamma_g\,,\hspace{0.5 cm} \Phi_f=F_f-c^{-1}\,\mathcal{H}_f\,\Gamma_f\,,\\
&\mathcal{E}=E_g-E_f\,,\hspace{0.5 cm} \mathcal{B}=B_f-c^2\,B_g+(1-c^2)E_g'\,,\nn
\ees
where $\Gamma_{g, f}\equiv B_{g, f}+E_{g, f}'$. In the vector sector we have 4 fields and 1 gauge freedom and hence we can form 3 gauge-invariant combinations which we choose as follows~\cite{Comelli:2012db,DeFelice:2014nja}
\be
V_{g,f i}=C_{g,f i}-\mathcal{V}'_{g,fi}\,,\hspace{0.5 cm} \chi_i=C_{g i}-C_{fi}\,.
\ee

The energy-momentum tensor for the perturbed universe is
\be
T^{\mu}_{\nu}=\bar{T}^{\mu}_{\nu}+\delta T^{\mu}_{\nu}\,.
\ee
The perturbations can be divided in perfect-fluid and non-perfect-fluid ones, with 5+5 dof (degrees of freedom). The perfect fluid dof in $\delta T_{\nu}^{\mu}$ are those which keep $T^{\mu}_{\nu}$ in the perfect fluid form:
\be
T_{\nu}^{\mu}=(p+\rho)\,u^{\mu} \,u_{\nu}+p\,\delta^{\mu}_{\nu}\,.
\ee
We suppose here that the perturbations are only of this type. Thus, they are given by the density perturbation, the pressure perturbation and the velocity perturbation. Explicitly: 
\be
p=\bar{p}+\delta p\,,\hspace{0.5 cm} \rho=\bar{\rho}+\delta \rho\,, \hspace{0.5 cm} u^i=\bar{u}^i+\delta u^i=\delta u^i\equiv \frac{1}{a}v_i\,.
\ee
The $\delta u^0$ is not an independent dof, it is fixed by  the normalisation  $u_{\mu}u_{\nu}\,g^{\mu\nu}=-1$. 

We can now write the perturbed Einstein equations for the two metrics. In the following we will use the Fourier transform of perturbations with respect to $x^i$, the corresponding 3-momentum will be $k^i$ and $k^2\equiv k_i k^i$. To keep the notation simple, the Fourier transform will be denoted by the same symbol as the original function. 

\section{Tensor perturbations}\label{tensors}

 Tensor perturbations of a given $\bk$-mode are composed of two independent helicity modes,
 \be
 h^{TT}_{ij} =  h^+e^{(+2)}_{ij}  + h^{-}e^{(-2)}_{ij}
 \ee
 where $+$ and $-$ denote the two helicity-2 modes of the gravitational wave. For an orthonormal system $(\widehat \bk,\bfe^{(1)},\bfe^{(2)})$ we have
 \be \bfe^{\pm} =\frac{1}{\sqrt{2}}\left(\bfe^{(1)}\pm i\bfe^{(2)}\right) \quad \mbox{ and }
 \quad e^{(+2)}_{ij} =\bfe^{+}_i\bfe^+_j \,, \quad  e^{(-2)}_{ij} =\bfe^{-}_i\bfe^-_j \,. 
 \ee
 
 In what follows we assume parity invariant perturbations, 
 $$\langle h^+(\bk)(h^+(\bk'))^*\rangle =k^3\langle h^-(\bk)(h^-(\bk'))^*\rangle = \delta(\bk-\bk')2\pi^2P_h(k)\,,$$
  and $\langle h^+h^-\rangle =0$. And we shall consider just one mode, say $ h_f^+=h_fG_f $ and $h_g^+=h_gG_g$, where $G_f$ and $G_g$ are uncorrelated Gaussian random variables with vanishing mean and variance 
  $\langle G_{g,f}(\bk)G_{g,f}(\bk')\rangle = \delta(\bk-\bk')2\pi^2$, so that $h_{g,f}$ is  the square root of the power spectrum. 
  All the following is also valid for the modes $h^-_{g,f}$ which are not correlated with  $h^+_{g,f}$ in the parity symmetric situation which we consider.
 
 With a perfect fluid source term, i.e. no anisotropic stress, in the first order modified Einstein equation, we obtain the following tensor perturbation equations
for our bimetric cosmology~\cite{Cusin:2014psa}.
 \be\label{e:hg}
h_g''+2\mathcal{H}\,h'_g+k^2 h_g+m^2a^2r\, \beta_1\left(h_g-h_f\right)=0\,,
\ee
\be\label{e:hf}
h_f''+\left[2\left(\mathcal{H}+\frac{r'}{r}\right)-\frac{c'}{c}\right]\,h_f'+c^2 k^2\,h_f-m^2\beta_1\frac{c\, a^2}{r}\, \left(h_g-h_f\right)=0\,.
\ee
In Ref.~\cite{Cusin:2014psa} we have solved these coupled differential equations in the radiation era and have found that $h_f$ has a growing mode, $h_f\propto \tau^3$ on large scales which via the coupling enhances also the mode $h_g$ of the physical metric. Here we solve these equations numerically and analytically in the inflationary regime, where sensible approximations can be introduced to simplify the system.  

\subsection{Analytical results during inflation}\label{anal}

Deep in the inflationary epoch, the potential $V(\phi)$ is very flat and the inflaton is slowly rolling. Since\, $p_{\phi}\simeq -\rho_{\phi}$,  it is legitimate to model this period as a  de\,Sitter phase with constant Hubble parameter $H=H_I\simeq$ const. (where the suffix 'I' hereafter stands for inflation). From eq.~(\ref{H}) it follows that during inflation $r=r_I=$ const., with $r_I^2\simeq 3 H_I^2/(m^2 \beta_4)\simeq  3 \left(H_I/H_0\right)^2$ and eq. (\ref{Bianchiconstraint})  gives for the lapse function of the $f$-metric $c\simeq$ const $\simeq 1$. With the parametrization $\HH=-1/\tau$ and $a=-1/(\tau\,H_I)$ (note that with this choice $\tau<0$ during inflation), and $m^2\beta_1a^2 \simeq (H_0/H_I)^2\tau^{-2}$.
eqs.~(\ref{e:hg}) and~ (\ref{e:hf})  can be approximated in a de\,Sitter universe as 
\be\label{e:hg2}
h_g''-\frac{2}{\tau}\,h'_g+k^2 h_g+\left(\frac{H_0}{H_I}\right)\frac{1}{\tau^2}\,\left(h_g-h_f\right)=0\,,
\ee
\be\label{e:hf2}
h_f''-\frac{2}{\tau}\,h_f'+k^2\,h_f-\left(\frac{H_0}{H_I}\right)^3\frac{1}{\tau^2}\,\left(h_g-h_f\right)=0\,.
\ee
These equations can be solved exactly in terms of oscillating and decaying modes.

We want to choose as initial conditions the quantum vacuum of the graviton degree of freedom. For tensor perturbations the canonically normalised  variables (recall that $M_f=M_g=M_p$) are given by 
\be
\left(Q_{g}\right)_{ij}=e_{ij}\,Q_{g}=M_p\,a\,\left(h_{ij}^{TT}\right)_{g}\,,\hspace{0.5 cm}\left(Q_{f}\right)_{ij}=e_{ij}\,Q_{f}=M_p\,b\,\left(h_{ij}^{TT}\right)_{f}\,.
\ee 
Equations (\ref{e:hg2}) and (\ref{e:hf2}) in terms of these new variables and recalling that $b= r\,a$ become
\be\label{PP1}
Q_{g}''+\left(k^2-\frac{2}{\tau^2}\right)\,Q_g+\left(\frac{H_0}{H_I}\right)\,\frac{1}{\tau^2}\,\left(Q_{g}-\frac{1}{r}Q_{f}\right)=0\,,
\ee
\be\label{PP2}
Q_{f}''+\left(k^2-\frac{2}{\tau^2}\right)\,Q_f-\left(\frac{H_0}{H_I}\right)^3\,\frac{1}{\tau^2}\,\left(r \,Q_{g}-Q_{f}\right)=0\,.
\ee
Since during inflation $r_I\simeq H_I/H_0\gg1$, eqs. (\ref{PP1}) and (\ref{PP2}) can be approximated as
\be\label{P1}
Q_{g}''+\left(k^2-\frac{2}{\tau^2}\right)\,Q_g+\left(\frac{H_0}{H_I}\right)\,\frac{1}{\tau^2}\,Q_{g}=0\,,
\ee
\be\label{P2}
Q_{f}''+\left(k^2-\frac{2}{\tau^2}\right)\,Q_f-\left(\frac{H_0}{H_I}\right)^2\,\frac{1}{\tau^2}\,Q_{g}=0\,.
\ee
For sub-horizon scales, $|k \tau|\gg 1$, eqs. (\ref{P1}) and (\ref{P2}) reduce to two copies of the same equation for a harmonic oscillator  with frequency $k$. 
The quantum vacuum solutions are
\be\label{initialcon}
Q_{g}=\frac{1}{\sqrt{2\,k}}\exp{\left(-i\,k\tau\right)}\,,\hspace{0.8 cm}Q_{f}=\frac{1}{\sqrt{2\,k}}\exp{\left(-i\,k\tau\right)}\,,\hspace{0.6 cm}\text{for}\qquad |k\tau|\gg1\,.
\ee

We want to solve eqs. (\ref{P1}) and (\ref{P2}) with initial conditions (\ref{initialcon}). These equations can be decoupled introducing the new variable $Q_{+}\equiv Q_f+\left(\frac{H_0}{H_I}\right)Q_g$
\be\label{PPP1}
Q_{g}''+\left(k^2-\frac{2}{\tau^2}\right)\,Q_g+\left(\frac{H_0}{H_I}\right)\,\frac{1}{\tau^2}\,Q_{g}=0\,,
\ee
\be\label{PPP2}
Q_{+}''+\left(k^2-\frac{2}{\tau^2}\right)\,Q_+=0\,.
\ee
Eqs. (\ref{PPP1}) and (\ref{PPP2}) can be solved in terms of Bessel functions.
Requiring that the asymptotic behavior (\ref{initialcon}) is recovered for $|k\,\tau|\gg 1$, we find the following solutions for the canonically normalized variables $Q_g$ and $Q_f$
\be\label{solQg}
Q_g=-\sqrt{\frac{\pi}{2}}\,\,\sqrt{\frac{k\,\tau}{2\,k}}\,J_{\frac{1}{2}\,\sqrt{9-4\gamma}}\,(k\tau)+i\,\sqrt{\frac{\pi}{2}}\,\,\sqrt{\frac{k\,\tau}{2\,k}}\,Y_{\frac{1}{2}\sqrt{9-4\gamma}}\,(k\tau)\,,
\ee
\be\label{solQf}
Q_f=\frac{1}{\sqrt{2\,k}}\,\left(1+\frac{H_0}{H_I}\right)\,\left(1-\frac{i}{k\,\tau}\right)\,e^{-i\,k\tau}-\left(\frac{H_0}{H_I}\right)\,Q_g\,.
\ee
To simplify the notation, we have introduced the tiny constant $\gamma\equiv H_0/H_I$.\footnote{Given that for $|k \tau|\gg1$, the behavior of the Bessel functions is $J_{\,\frac{1}{2}\sqrt{9-4\gamma}}\left(k\tau\right)\rightarrow -\sqrt{\frac{2}{\pi k\tau} }\cos(k\tau)$ and $Y_{\,\frac{1}{2}\sqrt{9-4\gamma}}\left(k\tau\right)\rightarrow -\sqrt{\frac{2}{\pi k\tau} }\sin(k\tau)$, the asymptotic behavior of (\ref{solQg}) and (\ref{solQf}) for $|k\tau|\gg1$ is exactly of the type (\ref{initialcon}).} In the limit $\ga\ra 0$ we have $Q_g=Q_f=Q_+$.

The canonically normalized variables $Q_{g}$ and $Q_{f}$ are connected to the power spectrum by
\be
P_{h_g}(k)=k^3\,|h_{ij\,g}^{TT}\,\,h_{g}^{ij\,TT}|=2\,\cdot\frac{\,k^3\,|Q_{g}|^2}{a^2\,M_p^2}\,,
\ee
\be
P_{h_f}(k)=k^3\,|h_{ij\,f}^{TT}\,\,h_{f}^{ij\,TT}|=\frac{2}{r_I^2}\,\cdot\frac{\,k^3\,|Q_{f}|^2}{a^2\,M_p^2}\,,
\ee
where the factor of 2 is due to the two helicity modes in each tensor sector. Hence from eqs. (\ref{solQg}) and (\ref{solQf}) we can find the solutions of eqs. (\ref{e:hg2}) and (\ref{e:hf2}) making use of the relations
\be\label{link}
h_{g}=\frac{1}{a\,M_p}\,k^{3/2}\,Q_{g}\,,\hspace{1.5 cm}h_{f}=\frac{1}{r_I}\,\frac{1}{a\,M_p}\,k^{3/2}\,Q_{f}\,.
\ee

\subsection{Numerical results during inflation and reheating}

\begin{figure}[ht!]
    \centering
     \subfigure[\label{hgk1010}]
     {\includegraphics[scale=0.35]{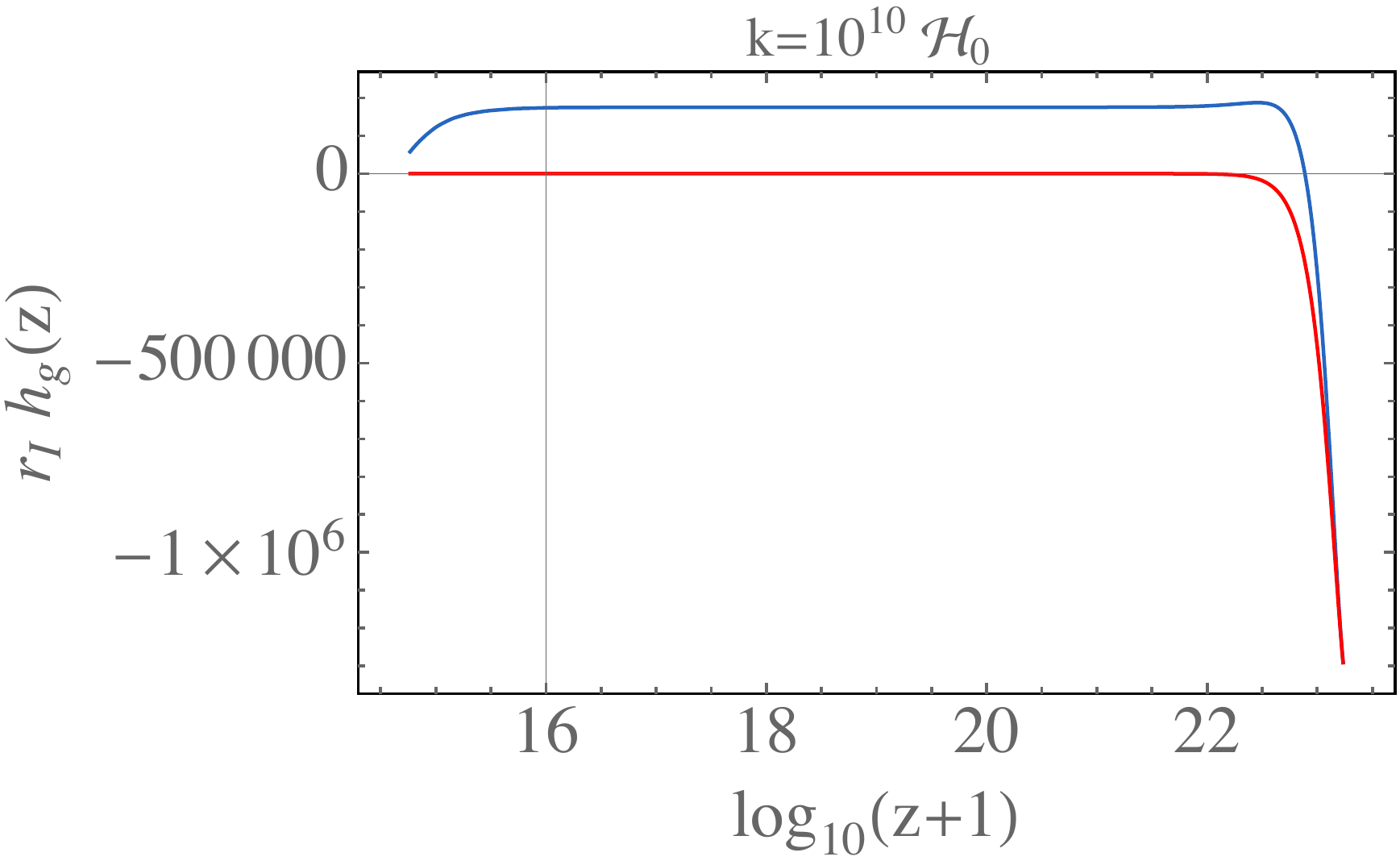}}\qquad\qquad\qquad
 \subfigure[\label{hfk1010}]
     {\includegraphics[scale=0.35]{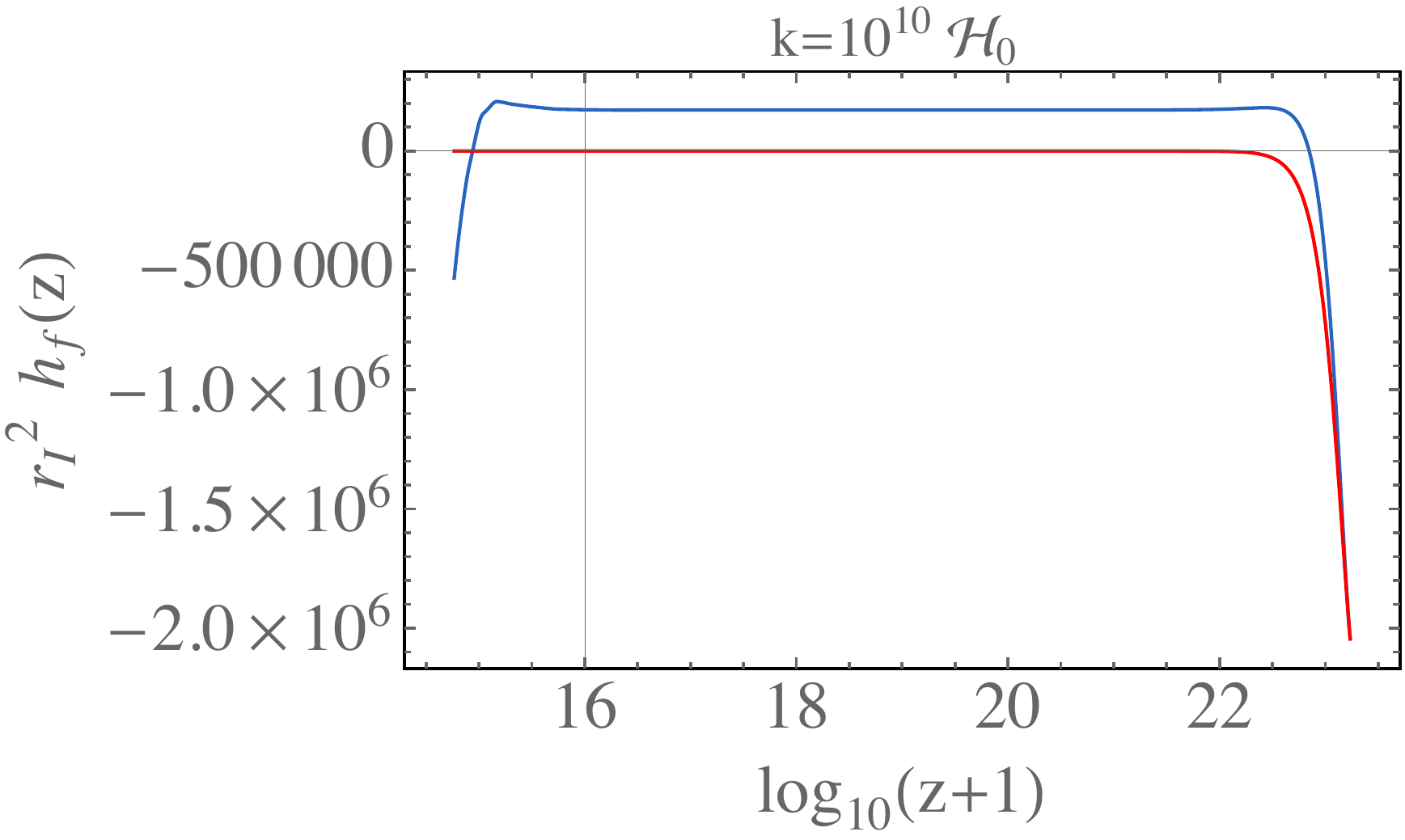}}
         \subfigure[\label{hgk1011}]
     {\includegraphics[scale=0.35]{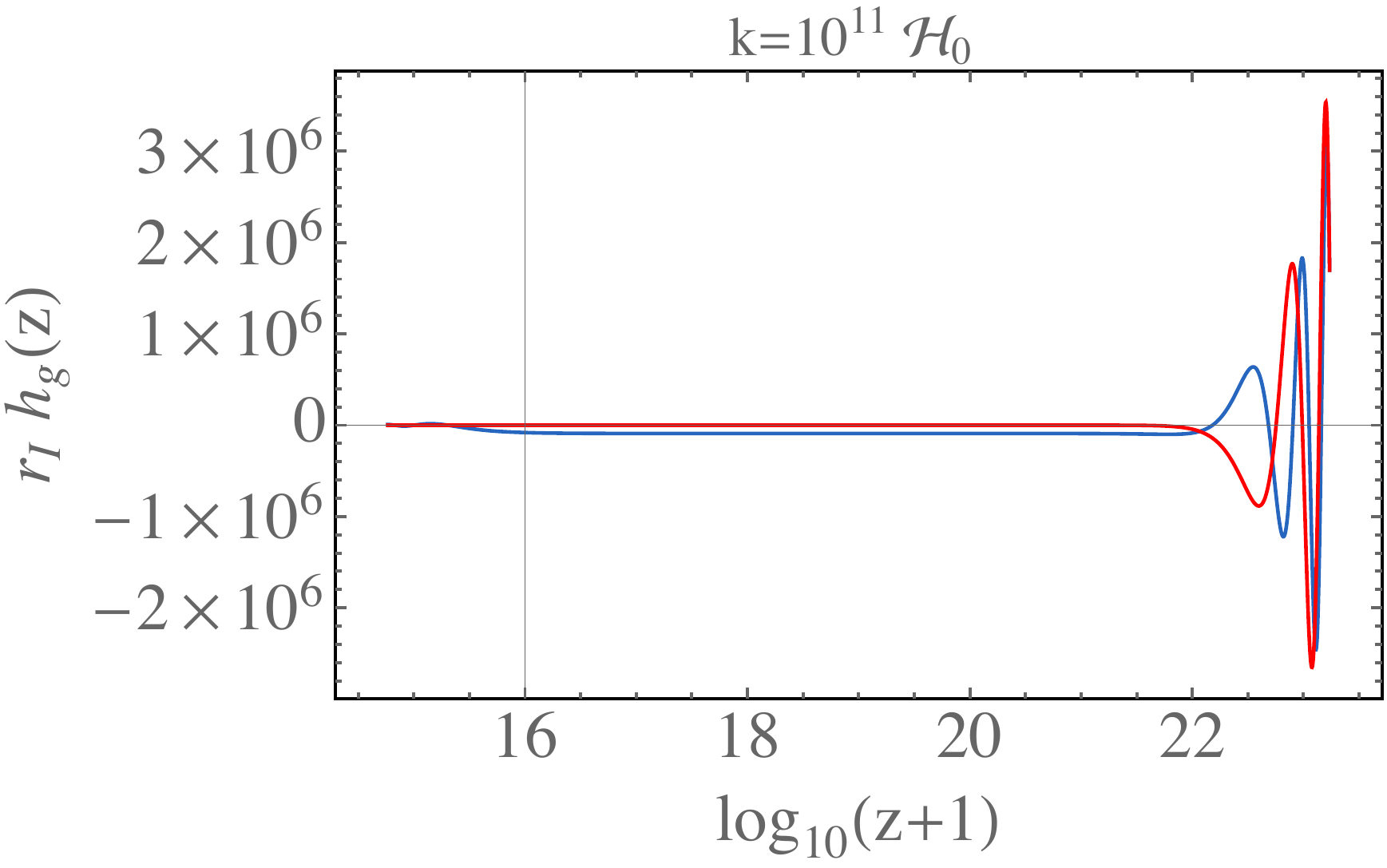}}\qquad\qquad\qquad
          \subfigure[\label{hfk1011}]
     {\includegraphics[scale=0.35]{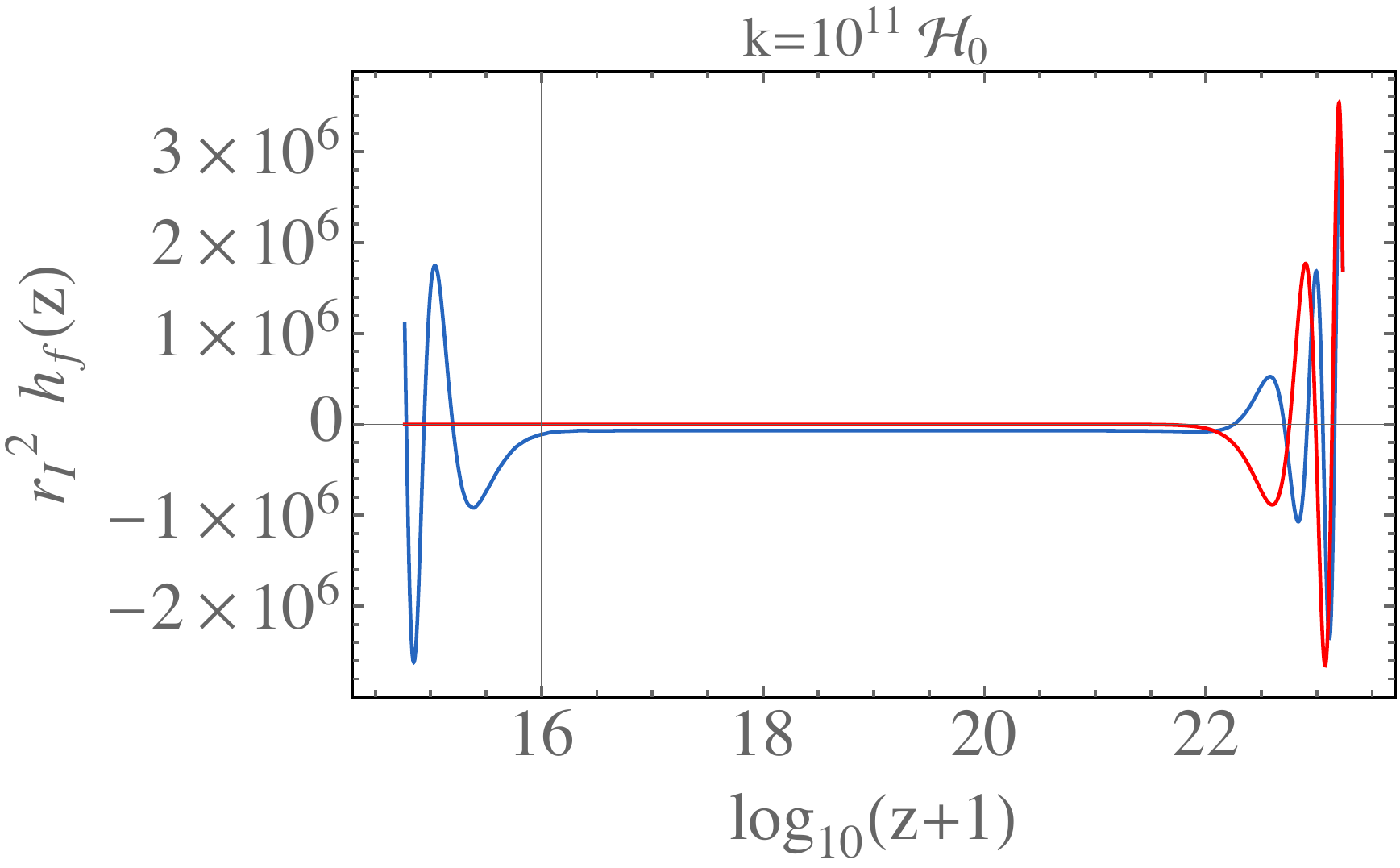}}
          \subfigure[\label{hgk1012}]
     {\includegraphics[scale=0.35]{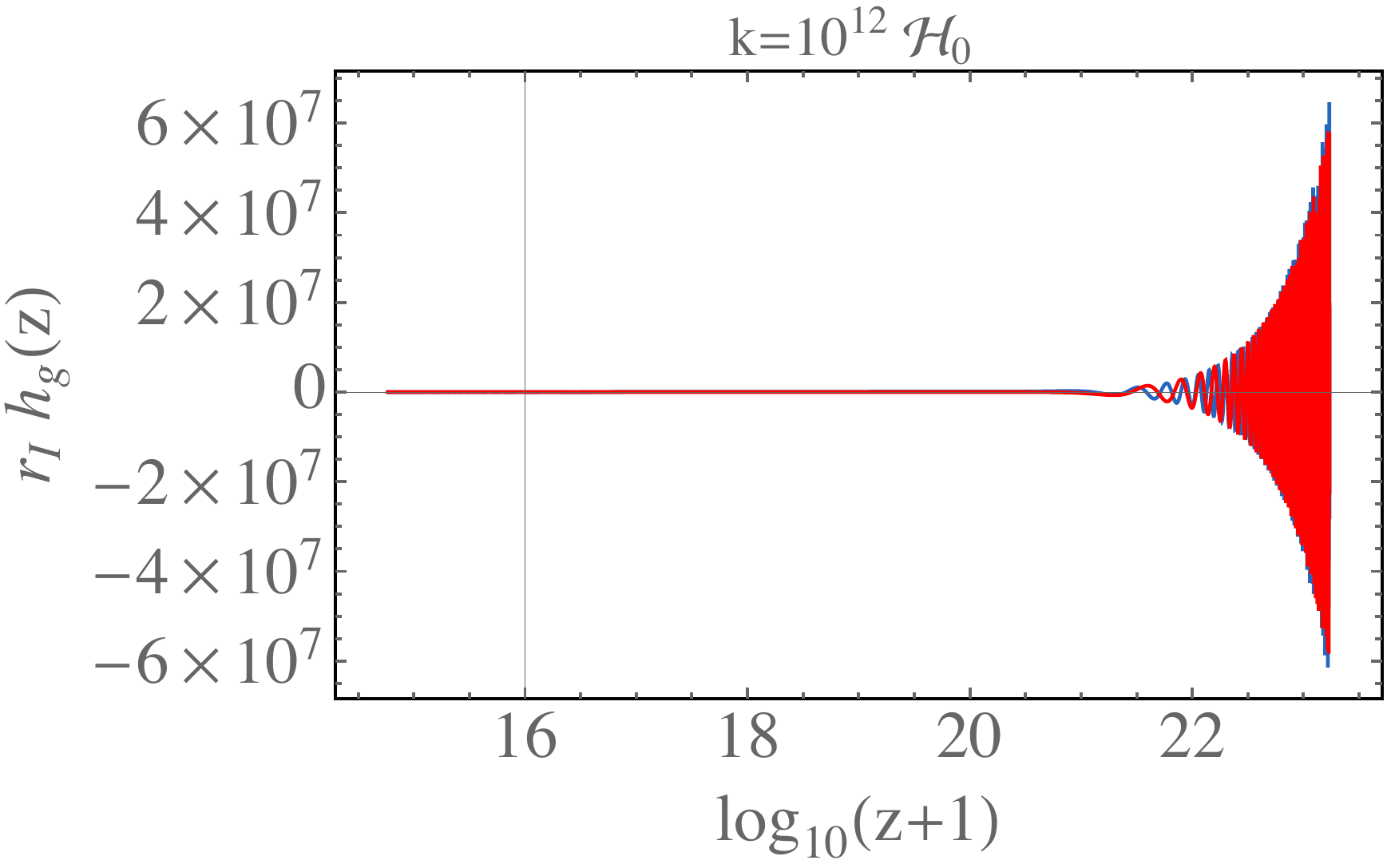}}\qquad\qquad\qquad
 \subfigure[\label{hfk1012}]
     {\includegraphics[scale=0.35]{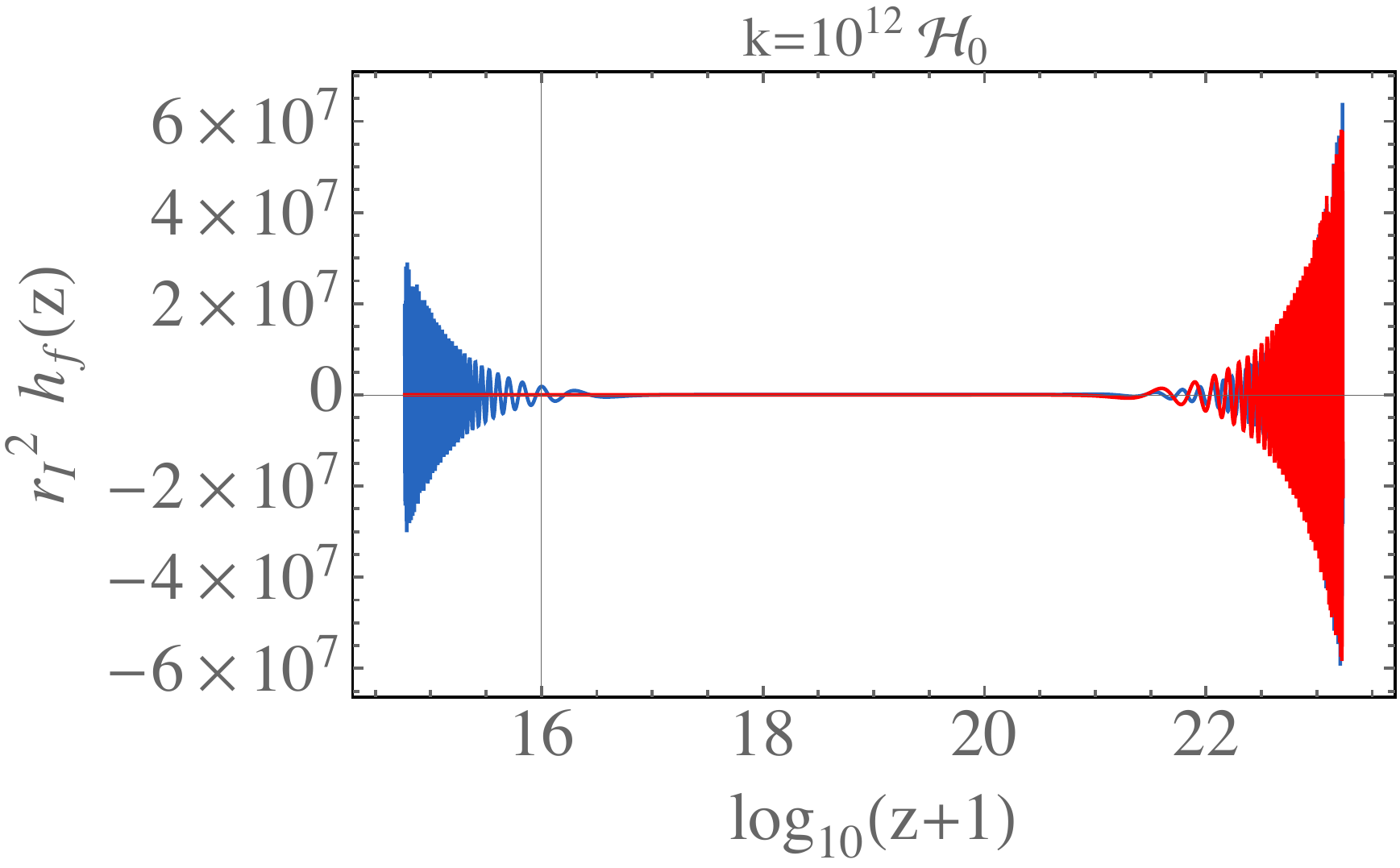}}
         \subfigure[\label{hgk1013}]
     {\includegraphics[scale=0.35]{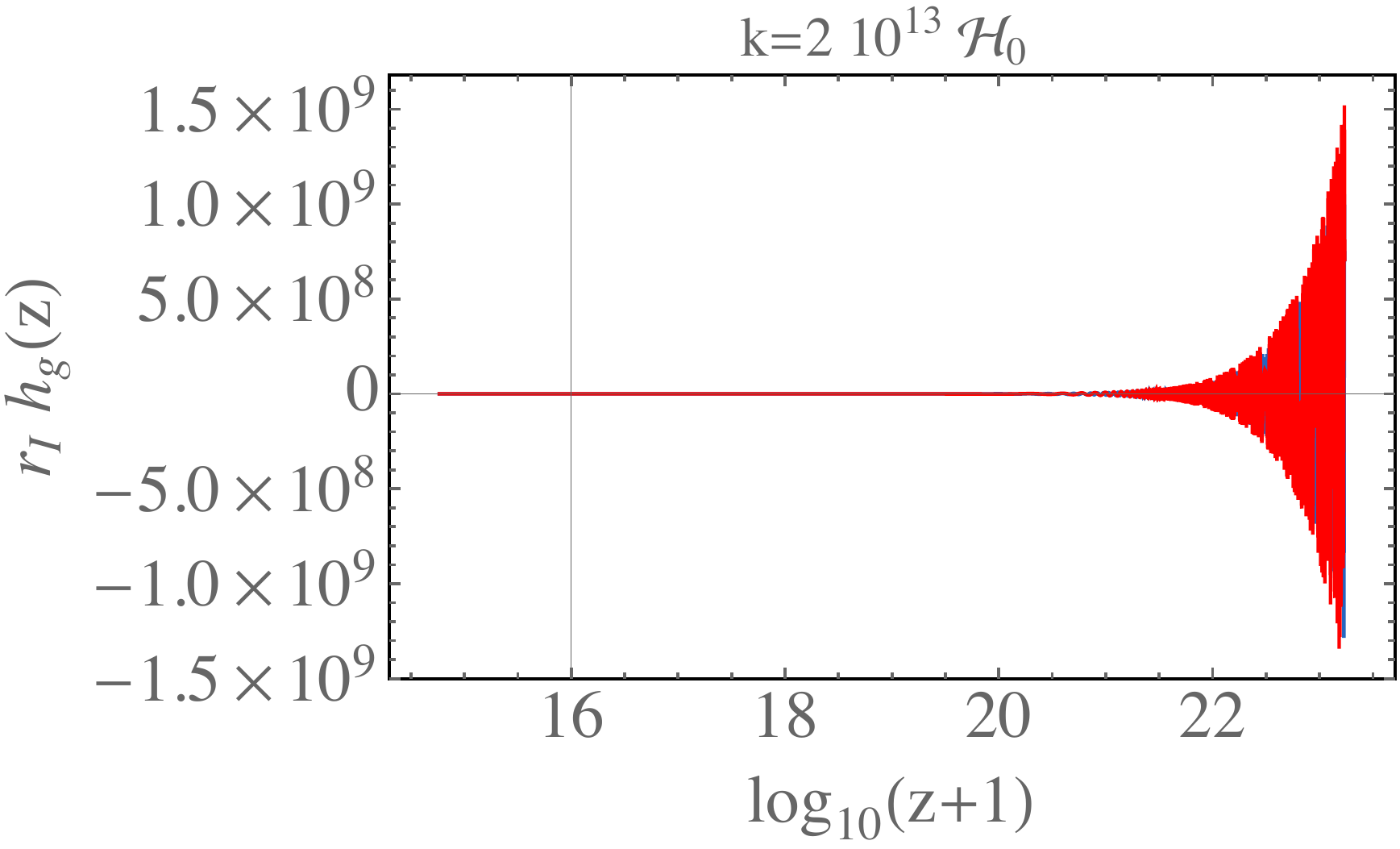}}\qquad\qquad\qquad
          \subfigure[\label{hfk1013}]
     {\includegraphics[scale=0.35]{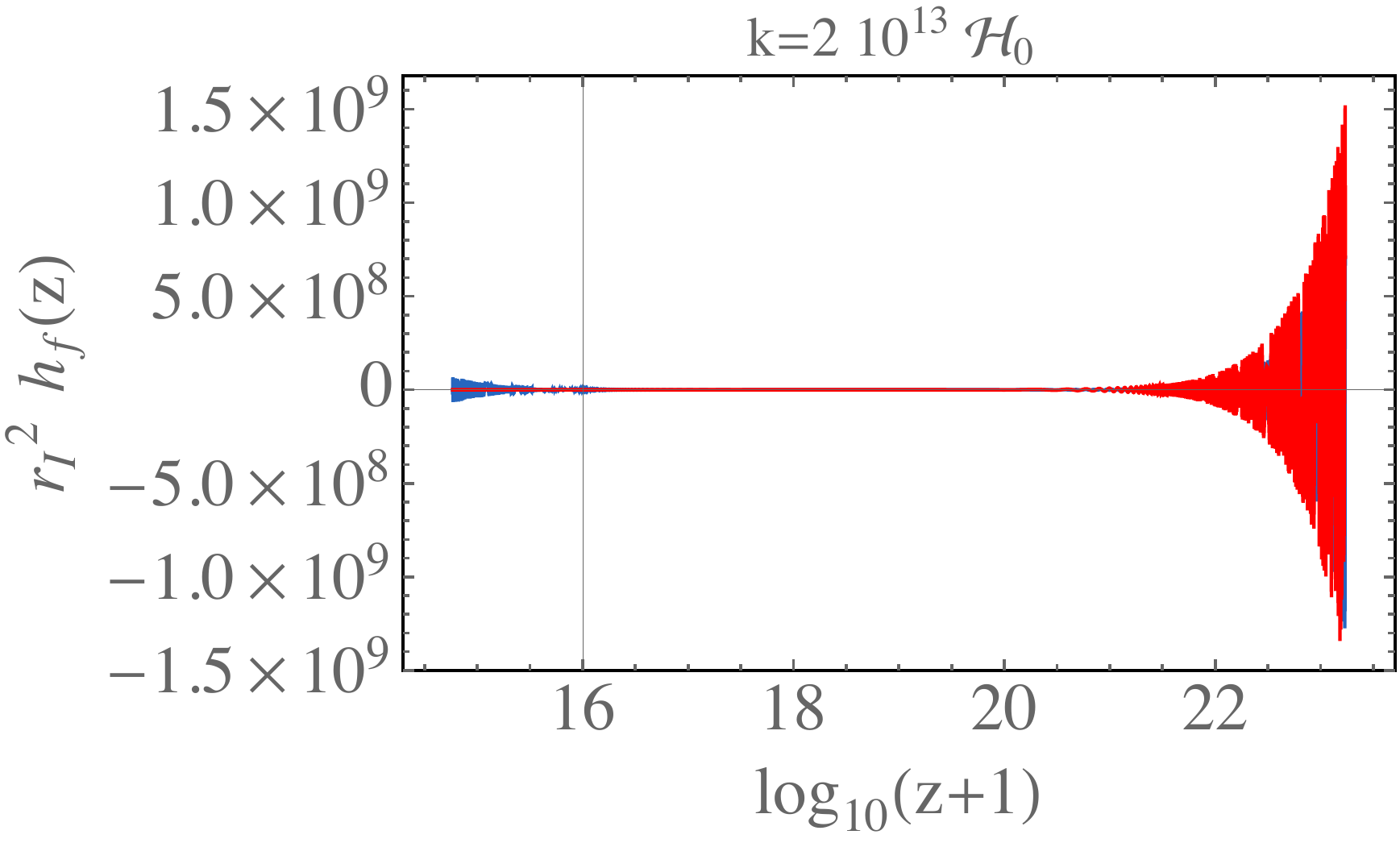}}
           \caption{\label{figpert1} Evolution of tensor perturbations of the metrics $g$ and $f$ as functions of redshift. The numerical solution (blue) is plotted together with the analytical one (in red) valid in the inflationary era, i.e., in the regime in which the hypothesis of slow-rolling holds. We have chosen $k=10^{10}\,\HH_0\,,$  $k=10^{11}\,\HH_0\,,$  $k=10^{12}\,\HH_0$ and  $k=2\cdot 10^{13}\,\HH_0$ in the panels \ref{hgk1010}-\ref{hfk1010}, \ref{hgk1011}-\ref{hfk1011}, \ref{hgk1012}-\ref{hfk1012} and \ref{hgk1013}-\ref{hfk1013}, respectively. The spectrum for the $g$-mode is rescaled with a factor $r_I\simeq10^{33}$ while the spectrum for the $f$-mode is rescaled with a factor $r_I^2$. Note that in our model inflation ends roughly at $\log_{10}(1+z)\simeq 19.0$ while radiation domination is established at $\log_{10}(1+z)\simeq 17.5$.}
 \end{figure}

\begin{figure}[ht!]
    \centering
     \subfigure[\label{loghgk1010}]
     {\includegraphics[scale=0.36]{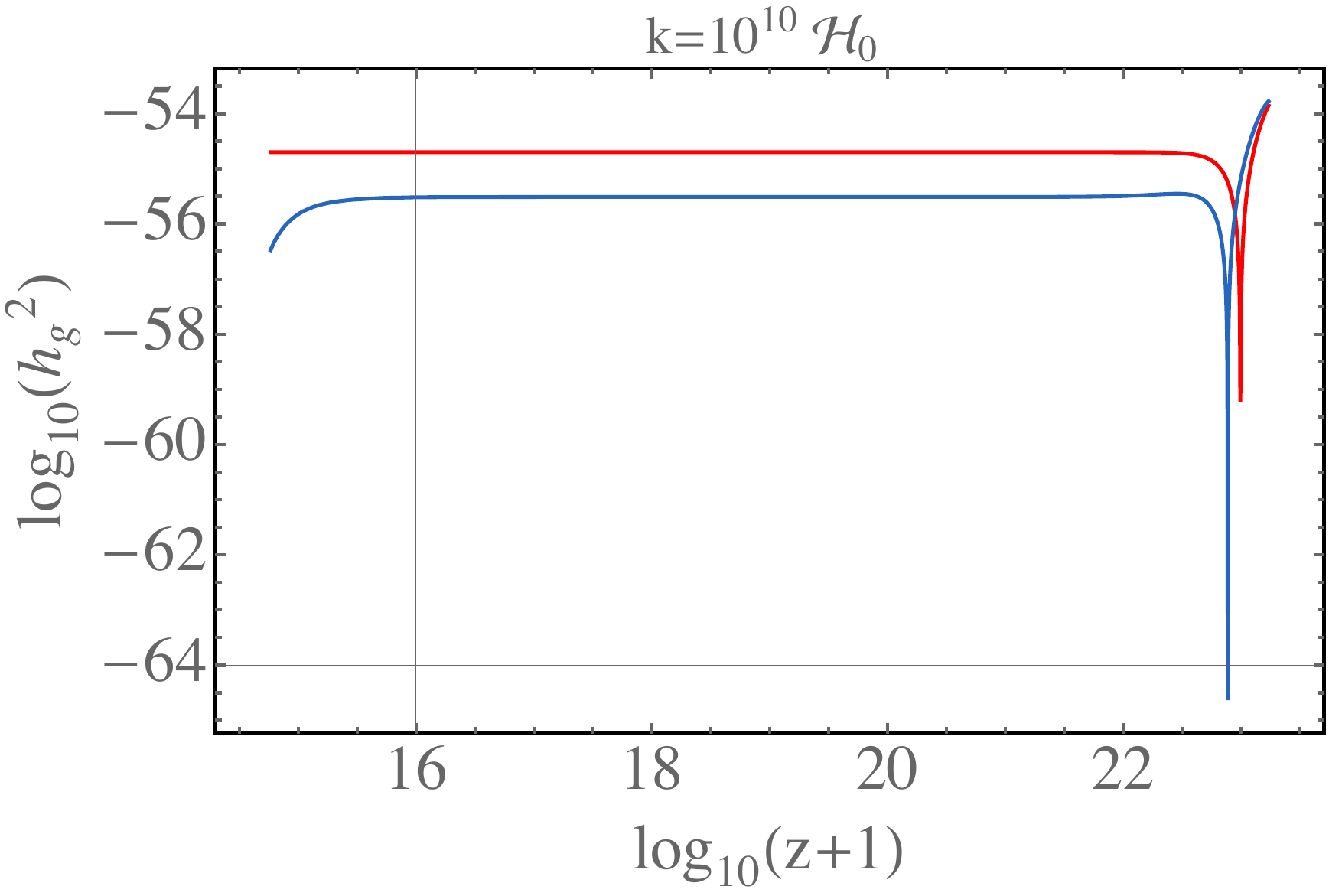}}\qquad\qquad\qquad
 \subfigure[\label{loghf1010}]
     {\includegraphics[scale=0.36]{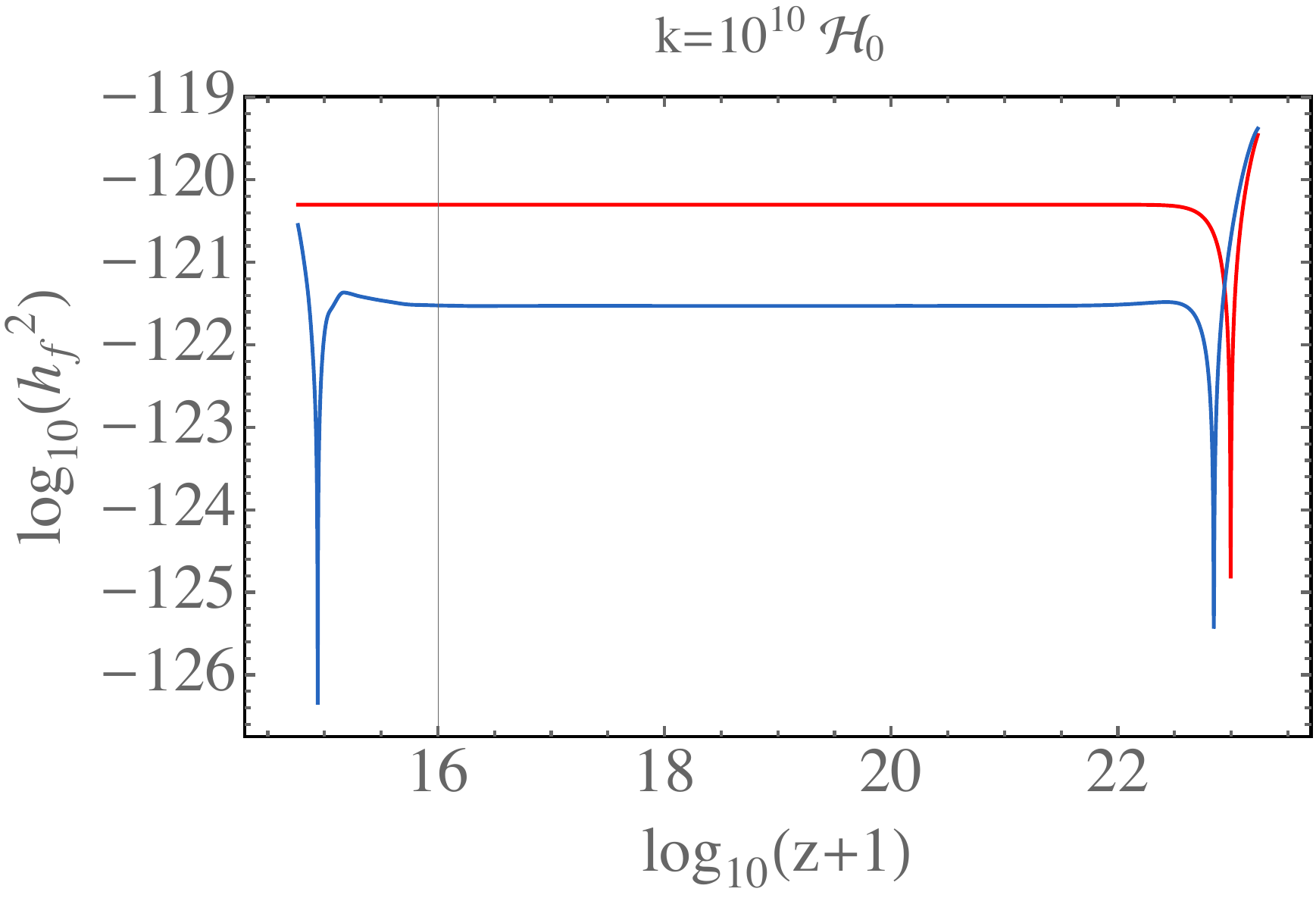}}
         \subfigure[\label{loghgk1011}]
     {\includegraphics[scale=0.36]{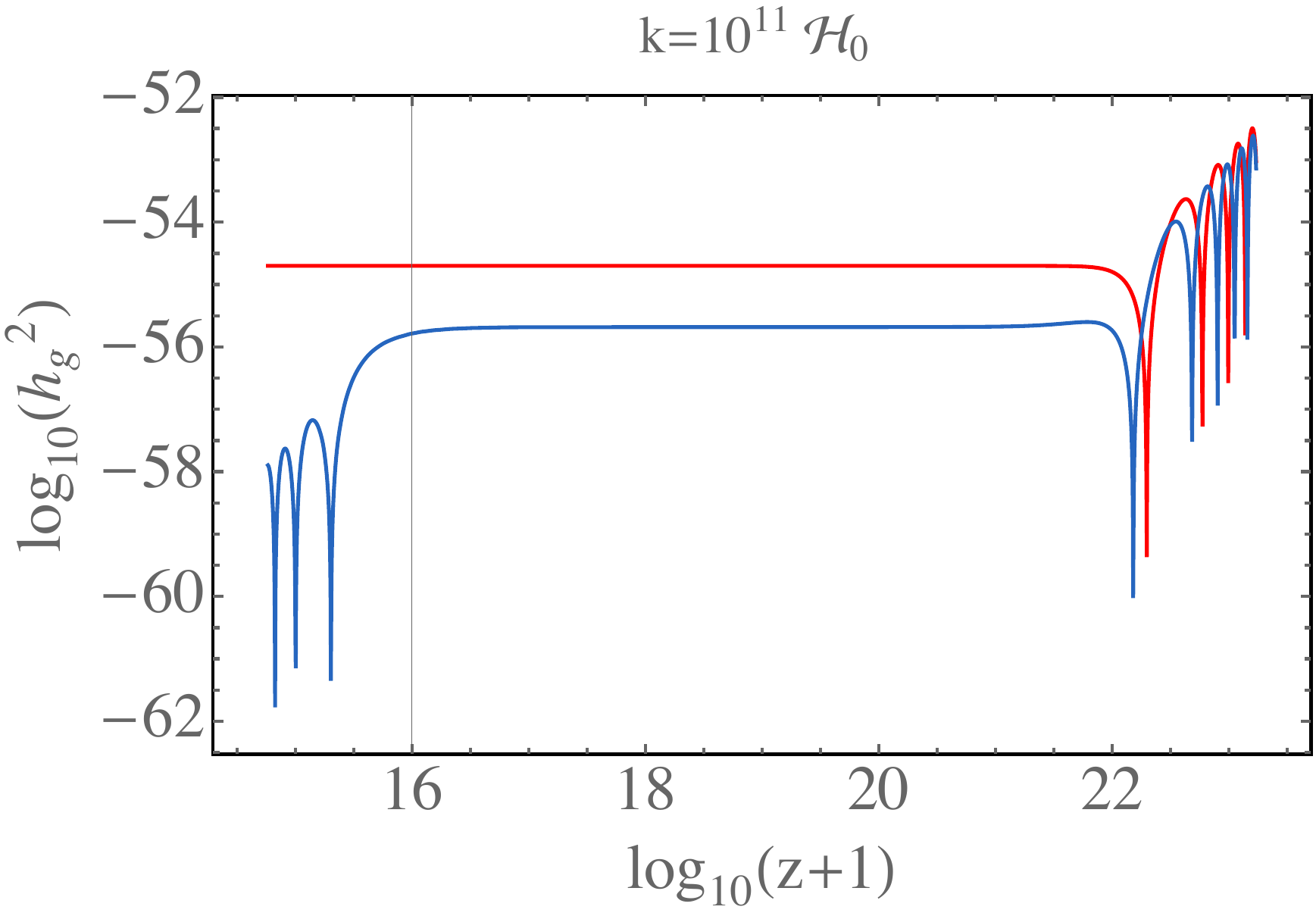}}\qquad\qquad\qquad
          \subfigure[\label{loghfk1011}]
     {\includegraphics[scale=0.36]{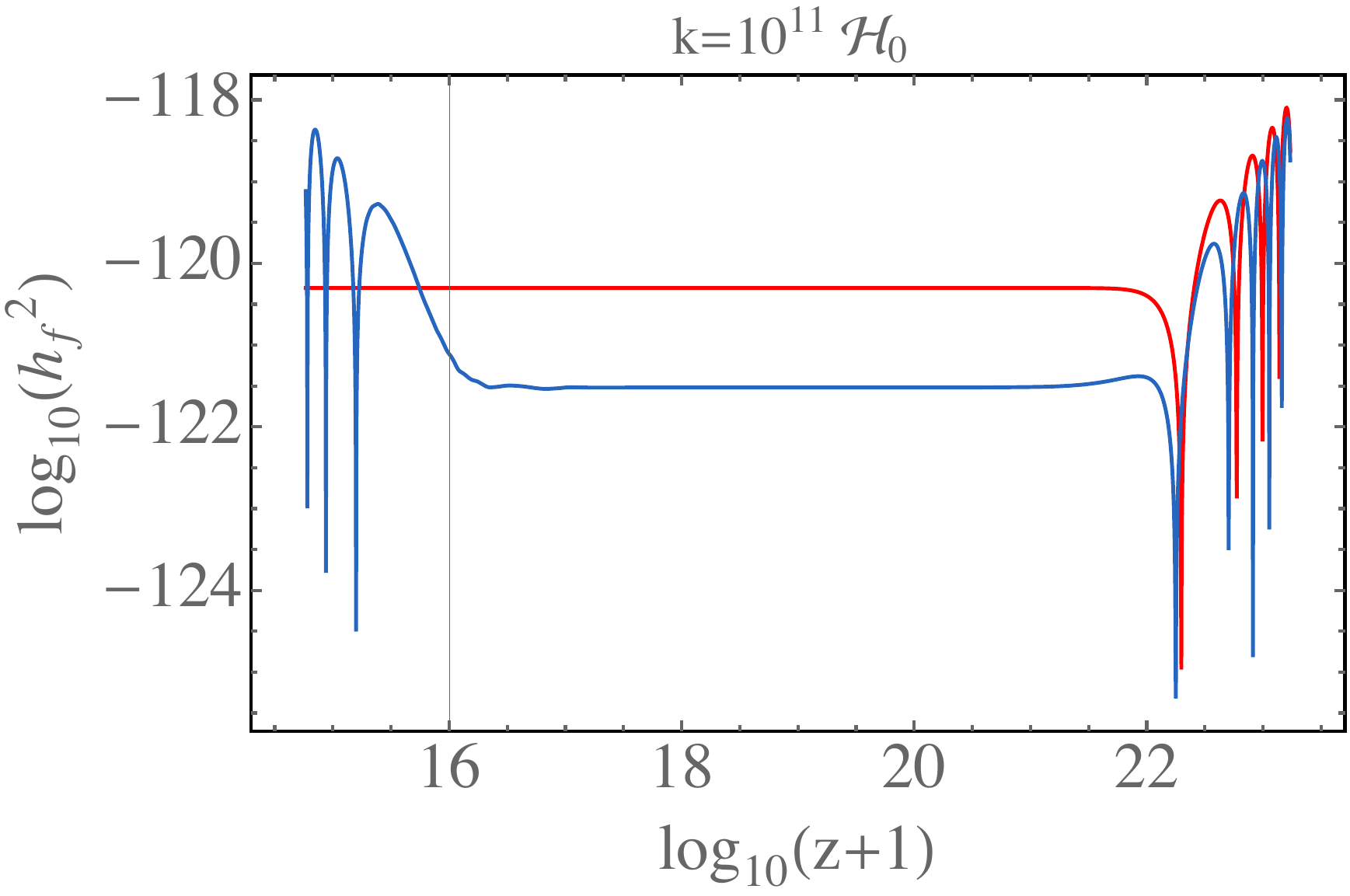}}
          \subfigure[\label{loghgk1012}]
     {\includegraphics[scale=0.36]{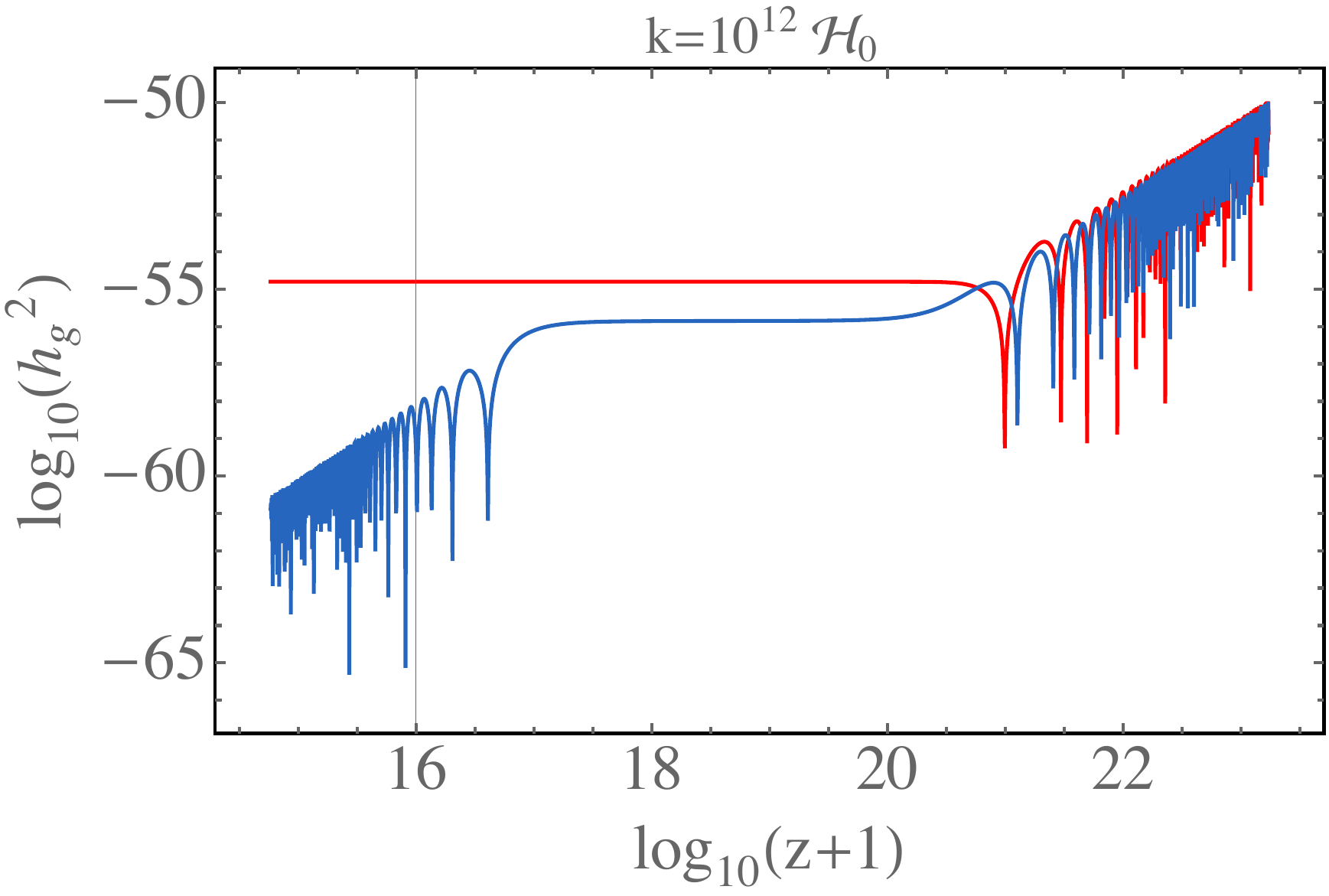}}\qquad\qquad\qquad
 \subfigure[\label{loghfk1012}]
     {\includegraphics[scale=0.36]{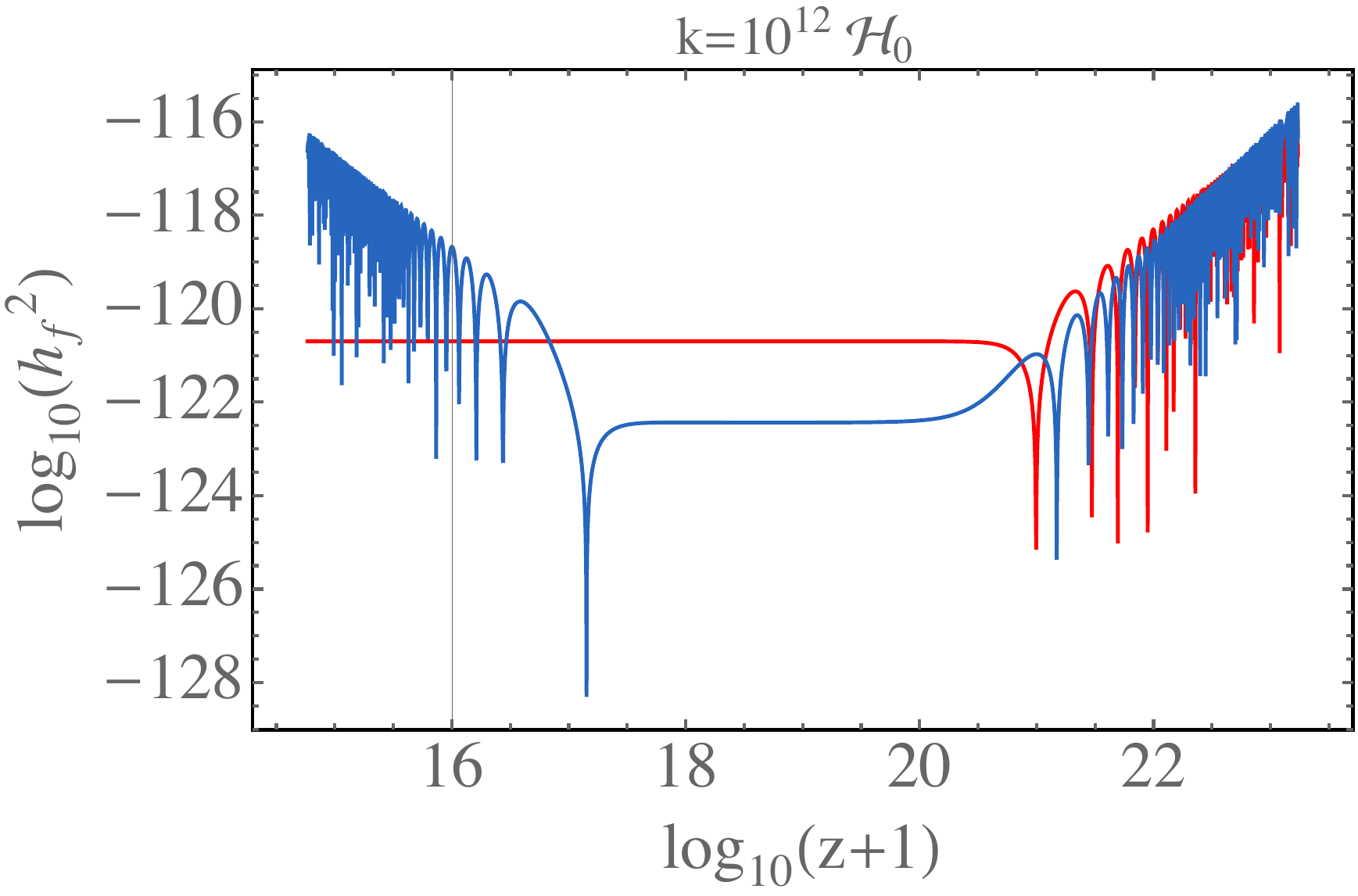}}
         \subfigure[\label{loghgk1013}]
     {\includegraphics[scale=0.36]{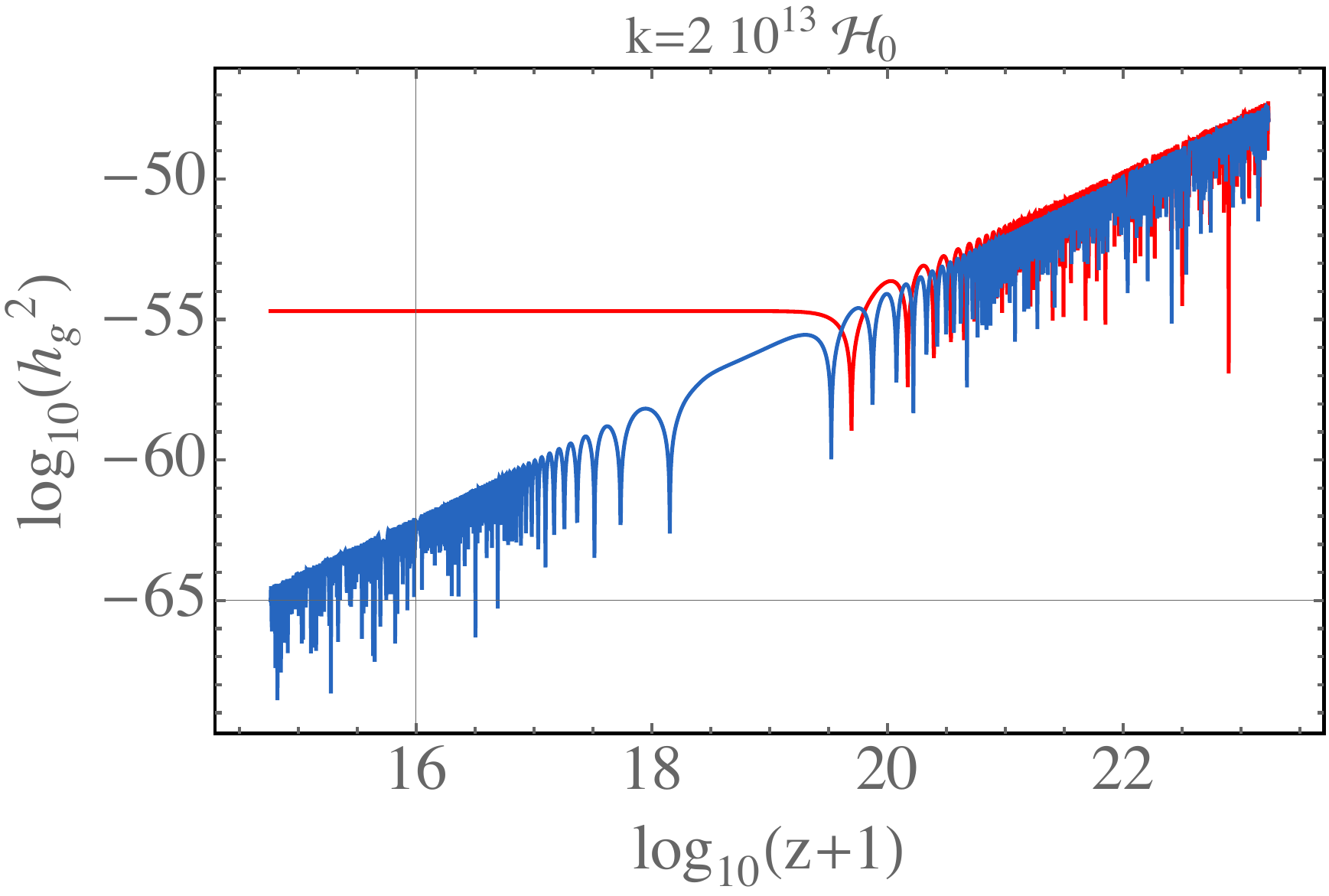}}\qquad\qquad\qquad
          \subfigure[\label{loghfk1013}]
     {\includegraphics[scale=0.36]{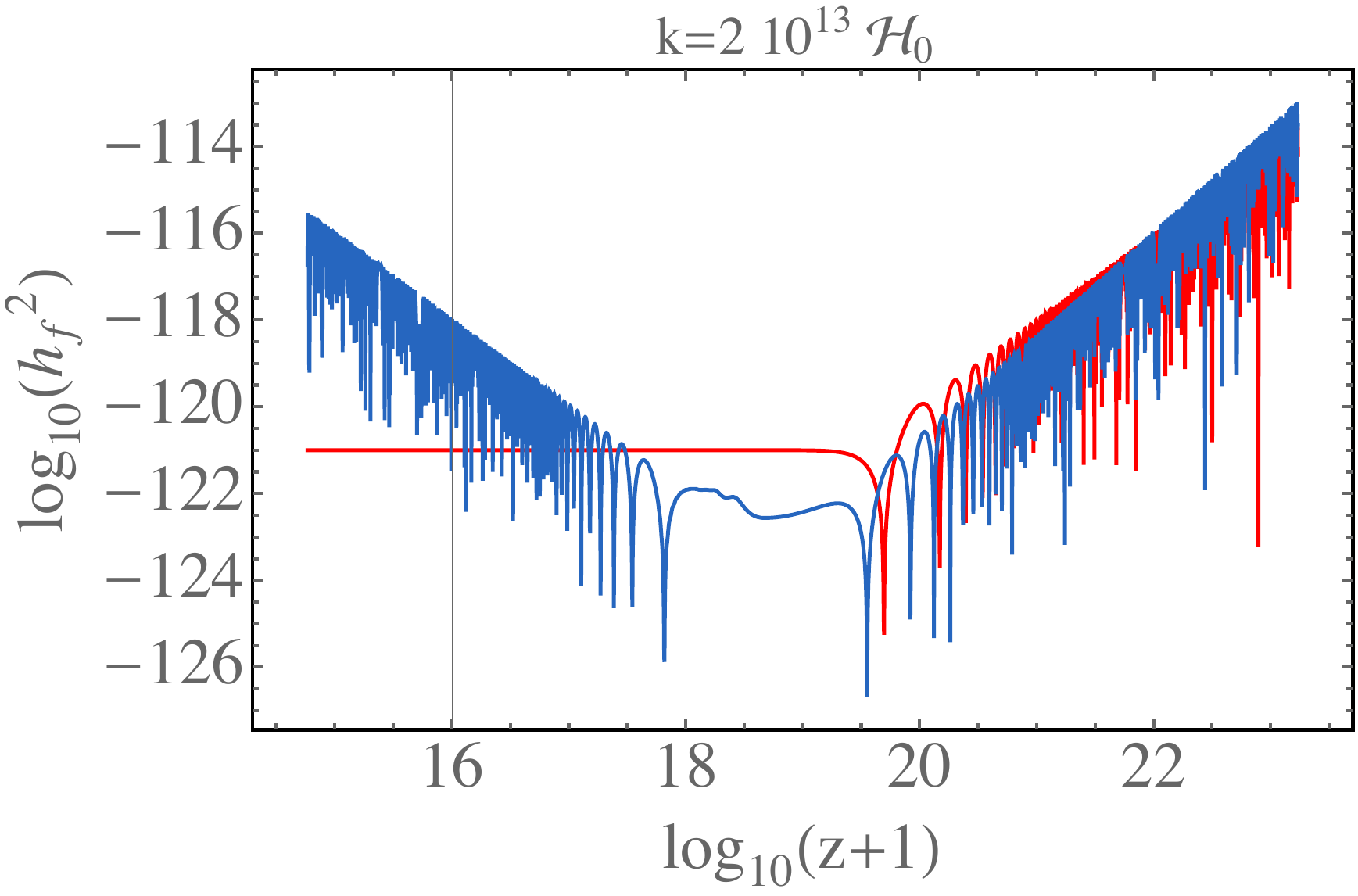}}
           \caption{\label{figpert2}  We show the same plots as in Fig.~\ref{figpert1} but in log scale for more detailed apprehension of the decay and growth of perturbations. One sees that the analytical solution during inflation is out of phase with the numerical solution. The reason is that the space-time background is somewhat different in the two cases, in fact, for our analytical solution the background is pure de\,Sitter, whereas for the numerical solution we have taken into account the full evolution of the background.} 
\end{figure}

The asymptotic behavior of the solutions (\ref{link}) for $|k\tau|\gg1$ during inflation is given by 
\be \label{hghf_asymp}
h_g=-\frac{k}{\sqrt{2}M_p}\,H_I\,\tau\,e^{-i\,k\tau}\,,\hspace{1 cm}
h_f=-\frac{k}{\sqrt{2}M_p}\,\frac{H_I\,\tau}{r_I}\,e^{-i\,k\tau}\,,\hspace{0.6 cm}\text{for}\qquad |k\tau|\gg1\,.
\ee
These functions  and their first derivatives can be evaluated at $\tau=\tau_{i}$ to find the initial conditions for the numerical evolution of  the full tensor perturbation equations, (\ref{e:hg}) and (\ref{e:hf}). The results of the numerical integration  are shown in Figs. \ref{figpert1} and \ref{figpert2}, for four different $k$-modes. 

Independently of the mode $k$, the agreement between the analytical solutions for $h_g$ and $h_f$, obtained from (\ref{solQg}) and (\ref{solQf}),  and the numerical one is reasonably good in the regime in which the slow-roll condition holds and the background is well approximated by de\,Sitter. As expected, both modes $h_g$ and $h_f$ oscillate in the redshift range in which $k>\HH(z)$, see eq. (\ref{hghf_asymp}). Furthermore, the mode $h_f$ develops an instability at the end of the inflationary period, where it oscillates with increasing amplitude. This is due to the fact that the damping term in eq. (\ref{e:hf}) becomes an anti-damping term at the end of the inflationary stage. Indeed, 
using eq.~(\ref{Bianchiconstraint}),  
eq.~(\ref{e:hf}) can be written as
\be\label{e:hf_bis}
h_f''+\left[2\,c\,\HH-\frac{c'}{c}\right]\,h_f'+c^2 k^2\,h_f-m^2\beta_1\frac{c\, a^2}{r}\, \left(h_g-h_f\right)=0\, .
\ee
Since $c=1$ at the beginning of the inflationary era whereas $c=-1$ in the radiation era, the term in square bracket changes sign when inflation ends, going from $2\HH$ to $-2\HH$, i.e.  from a positive damping term to a negative anti-damping term.

\begin{figure}[ht!]
    \centering
     \subfigure[\label{powerg}]
     {\includegraphics[scale=0.36]{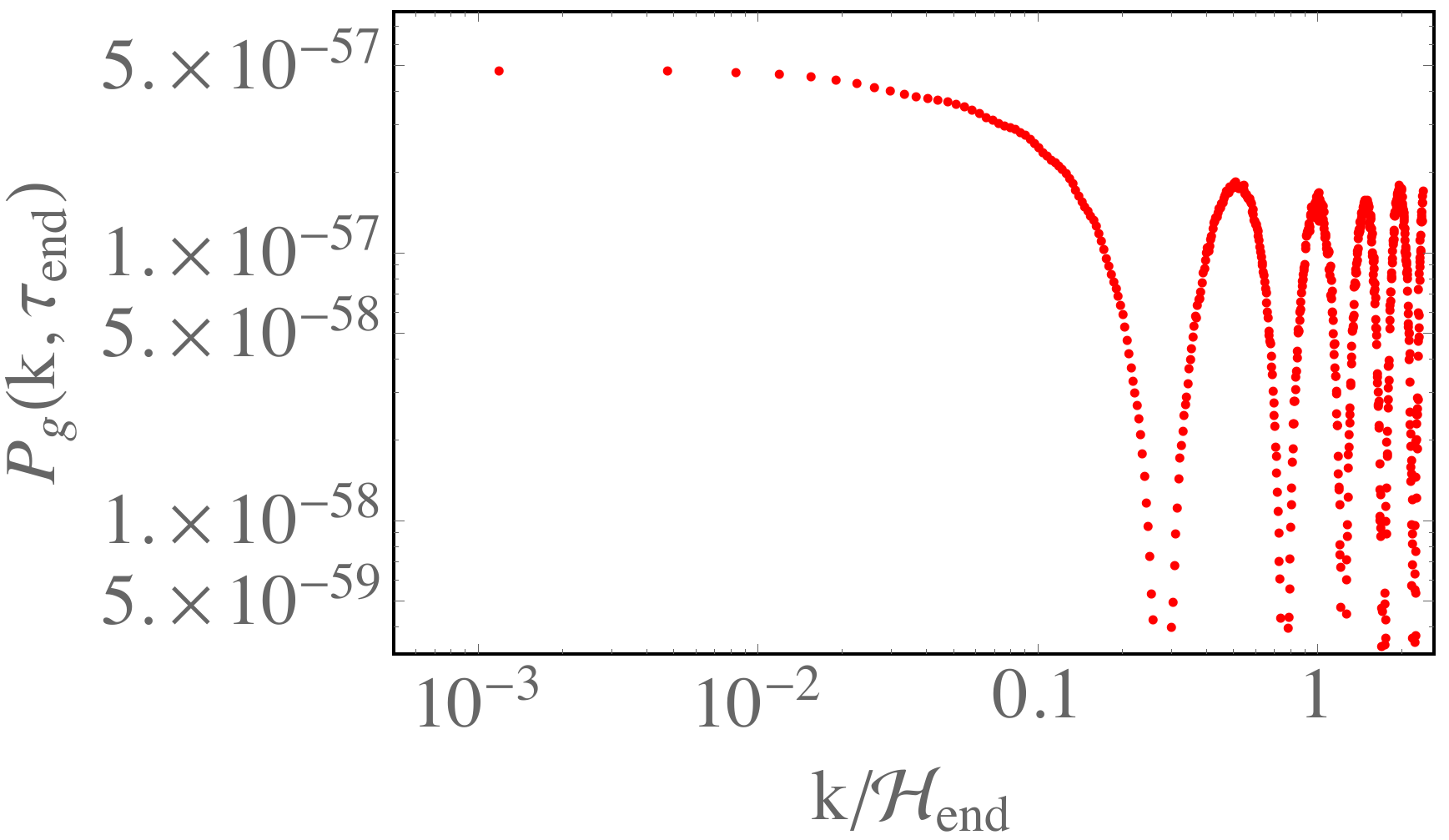}}\qquad\qquad\qquad
 \subfigure[\label{powerf}]
     {\includegraphics[scale=0.36]{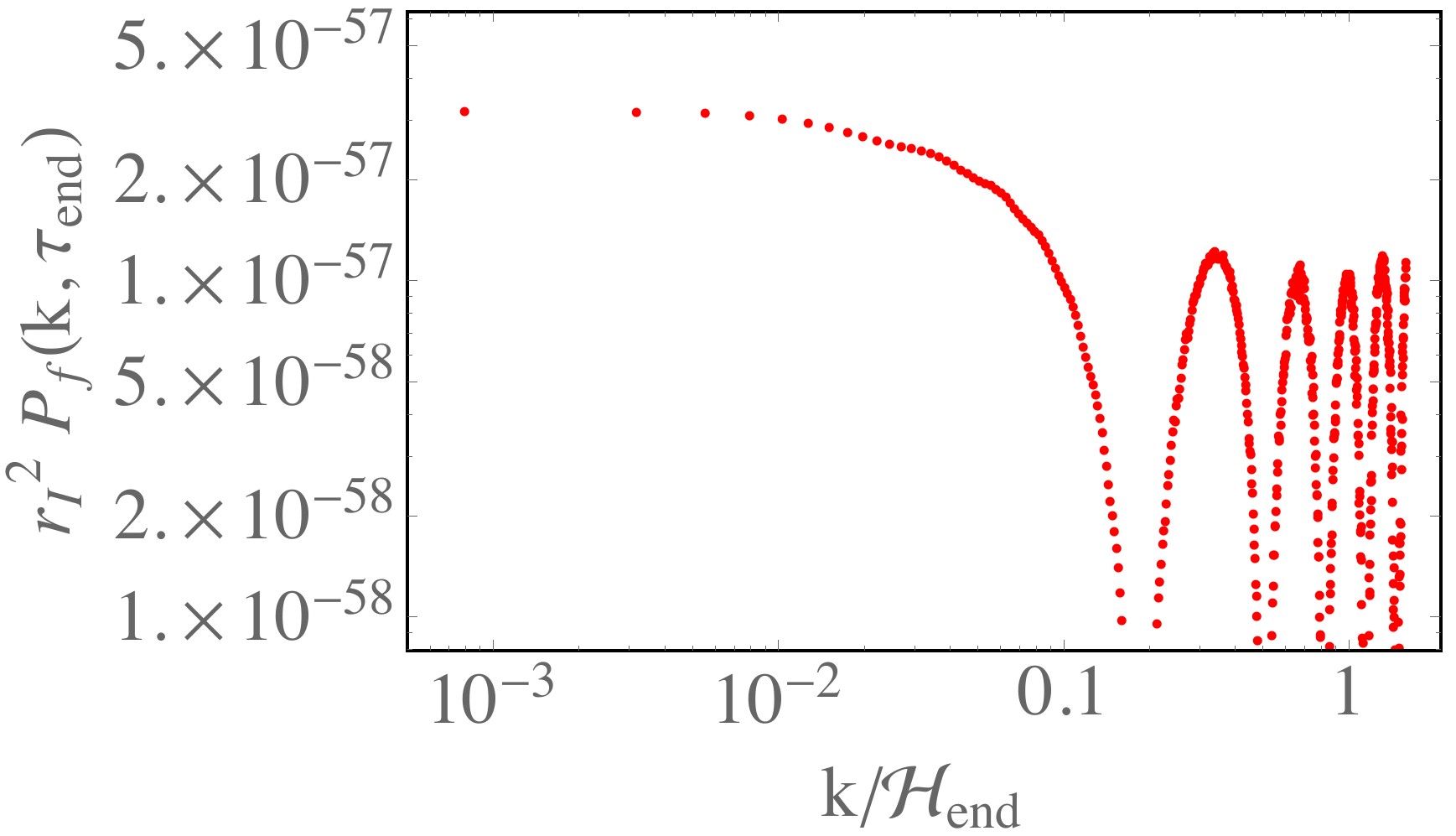}}
           \caption{\label{power} Power spectra at the end of inflation, $z\simeq 10^{19}$. The power spectrum for the $h_f$ mode has been rescaled by $r_I^2$, with $r_I\simeq 10^{33}$ for easier comparison with the spectrum of the physical mode $h_g$.} 
\end{figure}

From eqs. (\ref{solQg}) and (\ref{solQf}), it follows that on super horizon scales the power spectra at the end of inflation are scale invariant and given by
\be\label{e:gwspec}
P_{h_g}(z, k)\simeq\left(\frac{H_I}{M_p}\right)^2\simeq r_I^2\,P_{h_f}(z, k) \,,\hspace{2.5em} |k\,\tau| \ll1\,.
\ee
This result has also been derived in~\cite{Johnson:2015tfa}. $P_{h_g}$ hence has the same behaviour as the standard (i.e., GR) tensor power spectrum, whereas $P_{h_f}$ is suppressed with respect to $P_{h_g}$ by a huge factor $r_I^{2}$. 
The numerical results for the power spectra at the end of inflation are shown in Fig. \ref{power}.  In the analytical result (\ref{e:gwspec}) slow roll corrections are neglected since in this context we are mainly interested in orders of magnitude and not in very precise results. The power spectra shown in Fig.~\ref{power}, however, are numerically calculated with the full model.
 
 \section{Discussion: is the model still viable?}\label{discussion}

In \cite{Cusin:2014psa}, the cosmological evolution of tensor perturbations in the $\beta_1\beta_4$ model of bigravity has been addressed and the condition needed for the instability not to show-up until present times has been quantified in terms of a fine-tuning of the amplitude of the two tensor modes after inflation. 
We have found that during the radiation dominated era tensor fluctuations of the $f$ metric can grow like $a^3$ on super horizon scales. Furthermore, if they become larger than those of the $g$ metric, $h_f(t_0)>h_g(t_0)$,  they can influence the latter via the coupling term and show up e.g. in the CMB.

Considering naively, as from eq. (\ref{e:gwspec}), $h_f/h_g(z_e) = r^{-1}_I\simeq H_0/H_I$ at the end of inflation and that $h_f$ grows by an amount
$\left(T_{\rm eq}/T_{e}\right)^3 \simeq ((1+z_{\rm eq})/(1+z_e))^{3}$, the condition for the instability not to show-up $h_f(t_0)<h_g(t_0)$~\cite{Cusin:2014psa} implies a very low bound on the value of $H_I$ and therefore on the scale of inflation. Here the index $_e$ denotes the value of a quantity at the end of inflation.

However, this naive argument is misleading since at the end of inflation, when $|k\tau_e|\ll 1$ for the relevant modes,
\be
 h_f \sim r_I^{-1}\frac{H_I}{M_p}\left( 1 + |k\tau_e|^2\right) \,.
 \ee
Here, $\tau_e\simeq -(a_eH_I)^{-1}\simeq -(1+z_e)/H_I$ is the conformal time at the end of inflation. 
At the beginning of the radiation era the constant mode of  $h_f$ turns into the constant mode and only the very severely suppressed decaying mode, $\propto  |k\tau_e|^{2}$  turns into the growing mode. Therefore, considering only this decaying mode for $h_f$ at the end of inflation (when we put our initial conditions on the amplitudes of the tensor modes), our bound will be reduced by a factor $|k\tau_e|^{2} \simeq \left((1+z_e)k/H_I\right)^{2}$. This bound is valid for perturbations on very large scales with wave number $k\sim H_0$. Inserting this value for $k$, and replacing $\left((1+z_{\rm eq})/(1+z_e)\right)^3 \simeq (H_I/H_{\rm eq})^{3/2}$ by $(H_I/H_0)^{3/2}$ for simplicity\footnote{The last replacement is an approximation which corresponds to consider radiation domination lasting until present time.}, we obtain
\be
\left(\frac{h_f}{h_g}\right)(t_0) ~\lsim~   r_I^{-1}\left(\frac{H_0}{H_I}\right)^{-1/2} \simeq \left(\frac{H_0}{H_I}\right)^{1/2} \ll 1\,.
\ee
For the last  $\simeq$ sign we used that $r_I^{-1}\simeq H_0/H_I$. Therefore, no meaningful bound is obtained.  Also for the {smallest scales, with wave number} $k\simeq H_I/(1+z_e)$, there is no significant bound since inside the horizon the growth of $h_f$ is only linear in $a$ so that the factor  $\left(T_{\rm eq}/T_e\right)^3$ has to be replaced by $\left(T_{\rm eq}/T_e\right)$ and the same result is obtained. 

One may finally ask how model independent is the fact that the dominant constant mode {at the end of inflation} only transits into the constant mode of the radiation era. On super horizon scales and for negligible couplings the mode equations (\ref{e:hg}) and (\ref{e:hf}) are always of the form 
\be\label{e:gen}
h'' + \al(\tau)h' =0,
\ee
so they always have a constant mode. The general solution of (\ref{e:gen}) in fact is simply
\be
h(\tau) = A_1\int^\tau d\tau'\left[\exp\left(-\int^{\tau'}\al(\tau'')d\tau''\right)\right] +A_2 \,.
\ee
The first mode is decaying or growing, depending on the sign of $\al$ while the second mode is always constant. Hence even if there is a relatively long reheating phase, there will also be a constant mode during this phase.

Therefore, from a pure analysis of tensor perturbations, the model cannot be ruled out.

Note also that our inflationary model is typically above the strong coupling scale, $\La_3$, of the theory~\cite{ArkaniHamed:2002sp}.
For $T_{\rm reh}\simeq 10\,$~MeV we have $H_I\simeq T_{\rm reh}^2/M_P < 10^{-22}\,$~GeV $\simeq \La_3 = (m^2M_P)^{1/3} \sim 2\times 10^{-22}\,$~GeV. Therefore, if the reheating temperature is  above 10MeV, bigravity becomes strongly coupled and it is not granted that cosmological perturbation theory still applies.  However we underline that, since the strong coupling scale is derived in a Minkowski 
background, it is not entirely clear whether this represents  an upper limit on the Hubble scale or just on the energy of the perturbations on a given background. For the sake of the argument, our approach here is to simply analyse the classical bigravity Lagrangian in a Friedmann Universe with small perturbations.


\section{Vector perturbations}\label{s:vector}

Vector perturbations of a given $\bk$-mode can be decomposed as
\be
\mathcal{V}_{i}=\mathcal{V}^{(1)}e_{i}^{(1)}+\mathcal{V}^{(2)}e_{i}^{(2)}\, ,
\ee
where the two orthonormal vectors $e_{i}^{(1)}$ and $e_{i}^{(2)}$ are defined in Sec. \ref{tensors}. In what follows we shall consider just one mode, say  $\mathcal{V}^{(1)}$, since the situation is perfectly symmetric for the other mode and suppress the superscript, so that $\mathcal{V}_{i}\equiv e_i\,\mathcal{V}$. If the background is pure de\,Sitter, in the vector sector only the mode $\mathcal{V}_{i}\equiv \mathcal{V}_{gi}-\mathcal{V}_{fi}$ propagates~\cite{Comelli:2012db}. The action for this vector mode in Fourier space can then be written as \footnote{This action is equivalent to the action (63) in ref. \cite{DeFelice:2014nja}.}
\be\label{avec}
S_{\mathcal{V}}=\frac{M_p^2}{2}\int \mathrm{d}\tau \, \mathrm{d}^{3}k\,\frac{k^2\,a^4\,r\,m^2\,\beta_1}{k^2+a^2\,r\,m^2\,\beta_1}\,\left(| \mathcal{V}'_i |^2-\left(k^2+a^2\,r\,m^2\,\beta_1\right)\,|\mathcal{V}_i|^2\right)\, ,
\ee
where e.g. $|\mathcal{V}_i|^2\equiv \mathcal{V}_{i}^{*}\mathcal{V}^{i}$. The canonically normalized variable in this case is defined as
\be\label{rel}
\mathcal{Q}_i\equiv e_i\,\mathcal{Q}=M_p\,a^2\,k\,\sqrt{\frac{r\,m^2\,\beta_1}{k^2+a^2\,r\,m^2\,\beta_1}}\,\mathcal{V}_i\, .
\ee 
After integration by parts, the action (\ref{avec}) for the variable $\mathcal{Q}$ can be written as
\be\label{ap}
S_{\mathcal{V}}=\int d\tau \,  \mathrm{d}^{3}k\,\frac{1}{2}\left[\left(\mathcal{Q}'\right)^2-\,\mathcal{C}(k\,,\tau)\,\mathcal{Q}^2\right]\, ,
\ee
where, in order to simplify the notation, we have introduced
\be
\mathcal{C}(k\,,\tau)=k^{2}+\beta_1 \,m^{2}\, r\, a^{2}-2\,\mathcal{H}^{2}-\mathcal{H}'\biggl( \frac{k^{2}}{\beta_1 \,m^{2}\, r\, a^{2}+k^{2}} \biggr)-3\, \mathcal{H}^{2} \biggl( \frac{k^{2}}{\beta_1 \,m^{2}\, r\, a^{2}+k^{2}} \biggr)^{2}.
\ee
Using that $\beta_1m^2\sim H^2_{0}$ and that in pure de Sitter Universe with Hubble constant $H_I$, $a=-1/(H_I\tau)$ and $r_I\simeq H_I/H_0$, the previous expression can be approximated as
\be
\mathcal{C}(k\,,\tau)\simeq k^2+\left(\frac{H_0}{H_I}\right)\frac{1}{\tau^2}-\left(\frac{(k\tau)^2}{H_0/H_I+(k\tau)^2}\right)\,\frac{1}{\tau^2}-\left(\frac{\left(k\tau\right)^2}{H_0/H_I+\left(k\tau\right)^2}\right)^2\frac{3}{\tau^2}-\frac{2}{\tau^2}\simeq\, k^2-\frac{6}{\tau^2}\,,
\ee
where in the last equality we have used that $H_0/H_I\ll1$ and we have assumed that also $\beta_1m^2a^2r\ll k^2$ holds, or equivalently $H_0/H_I\ll (k\tau)^2$. This second inequality is valid for the following reason:
during the radiation era $ra^2=$const.$ \simeq \sqrt{3\Om_{\rm rad}}$ (This can be derived from the two Friedmann equations, see~\cite{Cusin:2014psa}.). Since this quantity is growing during inflation we have $r_Ia^2 \le \sqrt{3\Om_{\rm rad}}$.
Therefore $\beta_1m^2ra^2 \simeq H_0^2r_Ia^2\simeq (H_0/H_I)\tau^{-2} < H_0^2\sqrt{3\Om_{\rm rad}}<k^2$ for all values $k\gsim H_0$ which are observable.

The equation of motion derived from the action (\ref{ap}) with $\mathcal{C}(k\,,\tau)\simeq\, k^2-\frac{6}{\tau^2}$ is then \footnote{One can also verify that the exact equation of motion for $\mathcal{Q}$ which can be derived from the action (\ref{ap}) with the exact expression for $\mathcal{C}(k\,,\tau)$ coincides with eq. (7.9) in~\cite{Comelli:2012db}, once written in terms of the original variable $\mathcal{V}$.}
\be\label{eqq}
\mathcal{Q}''+\left(k^2-\frac{6}{\tau^2}\right)\,\mathcal{Q}=0\, .
\ee
For $|k\tau|\gg1$, Eq. (\ref{eqq}) reduces to a harmonic oscillator equation with frequency $k$, and has the vacuum solution
\be\label{asym}
\mathcal{Q}=\frac{1}{\sqrt{2 k}}\,e^{-i k\tau}\,,\hspace{1.5 cm} |k\tau|\gg1\,.
\ee
Eq. (\ref{eqq}) can be solved exactly. Asking that the asymptotic behavior (\ref{asym}) is recovered for $|k\tau |\gg1$ the solution is given by
\be\label{cr}
\mathcal{Q}= \frac{ik\tau}{\sqrt{2k}}h_2^{(2)}(k\tau)\,,
\ee
where $h^{(2)}_\ell$ is the spherical Hankel function of the second kind of order $\ell$, see~\cite{Abramo}. Substituting  eq. (\ref{cr}) in eq. (\ref{rel}), the evolution of the physical variable $\mathcal{V}$ can be written in terms of $Q$. Using again that $r_I\simeq H_I/H_0$, $\beta_1m^2a^2r\ll k^2$ and that $\beta_1m^{2}\sim H_0^{2}$, we obtain from eq. (\ref{rel})\be
\mathcal{V}_i\simeq \frac{1}{M_p}\,\frac{1}{a^2}\,\frac{1}{H_I}\,\sqrt{\frac{H_I}{H_0}}\,\mathcal{Q}_i\,.
\ee
For superhorizon modes $|k\tau|\ll1$, using the asymptotic behaviour of the spherical Hankel function, $h_2^{(2)}(x) \simeq -3/x^3$ for $|x|\ll 1$, we find
\be
\mathcal{V}_i\simeq -\frac{3}{\sqrt{2}}\,\frac{H_I}{M_p}\,\sqrt{\frac{H_I}{H_0}}\,k^{-5/2}\,e_i\,.
\ee
Therefore, for super-Hubble scales, the vector power spectrum can be written as
\be
P_{\mathcal{V}}(k\,, \tau)\equiv k^3|k\,\mathcal{V}_{i}|^2\simeq 2\cdot \frac{9}{2}\,\left(\frac{H_I}{M_p}\right)^2\,\frac{H_I}{H_0}\,,\hspace{1.2 cm}|k\tau|\ll1\, ,
\ee
where the multiplication by a factor 2 in the last expression is due to the two vector modes and the powers of $k$ are  introduced to make the power spectrum dimensionless. Note the large enhancement by a factor $H_I/H_0$ with respect to the standard tensor spectrum which is of order $(H_I/M_p)^2$. 

In order for  linear perturbation theory to remain viable, one has to request at least that $P_{\mathcal{V}}\,< \,1$, which means that our inflation model must be such that $H_I\,< \,10^{-2}$ GeV (where we have used the fact that $H_{0}\sim 10^{-42}$ GeV and that $M_p \sim 2.4\cdot 10^{18}$ GeV). This requires a rather low scale of inflation which is however acceptable. 

Asking that vector perturbation should not be larger than scalar perturbations after inflation, $P_{\mathcal{V}}\,< \,10^{-9}$~\cite{Ade:2015xua} would require  an inflationary Hubble scale of $H_I\,< \,10^{-5}$ GeV  which corresponds to a  reheating temperature of $T_{\rm reh} ~ \lsim ~ (H_IM_p)^{1/2} \sim 10^6$ GeV. This inflationary scale is not excluded, see~\cite{German:2001tz,Allahverdi:2006iq} but rather  low.  However, we have not studied the evolution of vector perturbation during the radiation era. If they decay, as in GR, this second limit is not relevant. It only applies if vector perturbations in bigravity stay constant during the radiation dominated Universe.  

\section{Scalar perturbations}\label{s:scalar}

Let us now turn to scalar perturbations. For this we assume that, like for the other degrees of freedom, the difference to GR during inflation mainly comes from the existence of additional modes, but that the GR-modes are not strongly affected, since the coupling between the GR-modes and the additional modes is suppressed by $H_0/H_I$. We therefore assume that the inflaton mode leads to a nearly scale invariant spectrum like in GR, and we only study the additional helicity-0 mode of the massive graviton.  For this we work in a pure  de Sitter background and neglect the slow roll and the inflaton perturbation. In this situation the helicity-0 mode of the massive graviton is the only dynamical scalar degree of freedom.
It is given by a linear combination of the two Bardeen potentials \cite{Comelli:2012db},
\begin{eqnarray}
\Phi & \equiv & \Phi_{g}-2r_I^{2}\Phi_{f}\,.
\end{eqnarray}
Its evolution is governed by the equation
\begin{eqnarray}
\Phi''+2{\cal H}\Phi'\left[\frac{2k^{4}}{9a^{2}{\cal H}^{2}m_{\Phi}^{2}+k^{4}-18{\cal H}^{4}}-1\right]  + && \nonumber \\
\frac{1}{3}\Phi\left[\frac{4\left(k^{6}-3k^{4}{\cal H}^{2}\right)}{9a^{2}{\cal H}^{2}m_{\Phi}^{2}+k^{4}-18{\cal H}^{4}}+3a^{2}m_{\Phi}^{2}-k^{2}-6{\cal H}^{2}\right] & = & 0,\label{eq:de sitter scalar}
\end{eqnarray}
where 
\be
m_{\Phi}^{2}=m^{2}\beta_{1}\left(\frac{1}{r_I^{2}}+1\right) \simeq m^2\beta_1 \sim H_0^2 \,.
\ee
Using the same approximations as for the vector mode, 
eq.~(\ref{eq:de sitter scalar})
can be approximated on sub-Hubble scales by
\begin{eqnarray}
\Phi''+2{\cal H}\Phi'+k^{2}\Phi=0 \,,\hspace{1.5em}|k\tau|\gg1 \,.\label{eq:canonical eq scalars}
\end{eqnarray}
Analogously to tensors, we quantize the scalar perturbations in order
to find the initial conditions. The canonical variable is given by
$\phi=M_p\,a\,\Phi$. In terms of this variable, eq. (\ref{eq:canonical eq scalars}) reduces to a  harmonic oscillator equation with vacuum solution $\phi(\tau)=e^{-ik\tau}/\sqrt{2k}$. It follows that 
\be
\Phi(\tau)=-\frac{H_I}{M_p}\,\frac{e^{-ik\tau}}{\sqrt{2k}}\tau\,,\hspace{1.5em}|k\tau|\gg1\,.
\ee
On super-Hubble scales eq. (\ref{eq:de sitter scalar}) can be approximated by
\be
\Phi''-2{\cal H}\Phi'-2{\cal H}^{2}\Phi=0\,,\hspace{1.5em} \left|k\tau\right|\ll1\,,
\ee
with general solution
\be
\Phi=c_{1}\tau+\frac{c_{2}}{\tau^{2}}\, , \hspace{1.5em} |k\tau|\ll1\,,
\ee
where $c_1$ and $c_2$ are integration constants, which can be fixed by matching the sub-Hubble solution and its derivative with the
one in the super-Hubble regime. 

Note that the mode $\propto c_2$ manifests an instability on super Hubble scales since $|\tau|$ is decreasing during inflation. This is the manifestation of the fact that also during inflation the Higuchi bound is violated in the scalar sector.  Indeed, for the scalar sector  (helicity-0 mode) the Higuchi bound of the $\beta_1\beta_4$ model is given by~\cite{Fasiello:2013woa,Lagos:2014lca}
$$ \beta_1\left(3r+\frac{1}{r}\right)  -2\beta_4r^2>0 \,,$$
which is violated for  $r> 1.02$. In our treatment, neglecting couplings of the scalar mode to other modes, the instability coming from this violation only shows up as a growth of perturbations on super-Hubble scales which leads to a red spectrum as we now show. Note also that the growth is exponential in physical time since $\tau^{-2} \propto a^2 \propto \exp(2H_It)$.

Working out the matching conditions explicitly we obtain 
\be
\Phi=-e^{i}\,\frac{H_I}{M_p}\,\frac{1}{\sqrt{2k}}\left(\left(\frac{i}{3}+1\right)\tau+\frac{i}{3}\,\frac{1}{\tau^2\,k^3}\right)\,,\hspace{1.5 em} |k\tau|\ll1\,,
\ee
The power spectrum for scalar perturbations on super-horizon scales can then be expressed as 
\be
P_{\Phi}(\tau,k)=k^3|\Phi|^2\simeq \frac{1}{18}\left(\frac{H_I}{M_p}\right)^2\left(\frac{\mathcal{H}}{k}\right)^4\propto k^{-4}\,,\hspace{1.5 em} |k\tau|\ll1\,.
\ee

This very red power spectrum is strongly enhanced on large scales, $|k\tau|\ll1$. Comparing it to the standard inflationary scalar power spectrum which is of the order of~\cite{Ade:2015xua} 
$$  P_s(z,k)\simeq \left(\frac{H_I}{M_{p}}\right)^2\frac{1}{\ep}\simeq 2\cdot 10^{-9}\,,$$
where $\ep<1$ denotes the slow roll parameter, one must conclude that this mode, if it transits to the radiation era completely spoils the observed large scale structure. However, for scalar perturbations the matching from inflation to the radiation era has to be studied carefully, it can even lead to a change in the power spectrum as found, e.g.,  for the inflationary magnetic mode studied in Ref.~\cite{Bonvin:2011dt}.  For this reason, we shall not draw strong conclusions from this result. 

Nevertheless, we request that $P_{\Phi}(z,k)<1$ for perturbation theory to remain valid during inflation, so that we can neglect back-reaction of the perturbation to the cosmic evolution.
At the end of inflation we have $r_Ia_{\rm end}^2 \simeq (ra^2)_{\rm rad} \simeq\sqrt{3\Om_r}$ so that $\HH_{\rm end}=|\tau_{\rm end}|^{-1}=H_Ia_{\rm end} \sim (H_IH_0)^{1/2}(3\Om_r)^{1/4}$. Inserting $k\sim H_0$ in $(\HH_{\rm end}/k)^{4}$ we obtain $(\HH_{\rm end}/H_0)^{4} \sim 3\Om_r(H_I/H_0)^2$ which leads to 
\be
P_{\Phi}(\tau_{\rm end},H_0)\simeq \frac{3\Om_r}{18}\left(\frac{H_I^2}{M_pH_0}\right)^2\,.
\ee
The condition  $P_{\Phi}(\tau_{\rm end},H_0)<1$ then becomes
\be  H_I< \left[\frac{18M_P^2H_0^2}{3\Om_r}\right]^{1/4} \sim 10^{-11}{\rm \,GeV}\,, \qquad
  V_I^{1/4}  \sim  (H_IM_P)^{1/2}\, \lsim \,  10^{4}{\rm \,GeV} \,.\ee
 Also this is indeed a rather low inflation scale. Requesting $P_\Phi<10^{-9}$ would reduce it by another two orders of magnitude.

\section{Conclusions}\label{conclusion}
In this paper we have studied the generation of  perturbations during inflation in a bimetric theory of gravity. We have analysed the evolution of the two tensor modes and we have found that both acquire a scale invariant spectrum with $h_f=h_g/r_I$, where the ratio $r_I=\left.(b/a)\right|_I \simeq H_I/H_0\gg 1$ is nearly constant during inflation. In addition to  this constant tensor mode, the $f$-metric sector has also a decaying mode, which is the one that turns into the growing mode during the subsequent radiation era. Nevertheless, this mode is so severely suppressed that it does not lead to a significant amplification  of the physical tensor mode as discussed in Ref.~\cite{Cusin:2014psa}.  Therefore, looking at the tensor sector alone, the $\beta_1\beta_4$ model of bimetric gravity cannot be ruled out despite the fact that the Higuchi bound of the tensor sector of the $f$-metric is violated. Note that this Higuchi bound on a Friedmann background does not lead to an exponential instability but only to power law growth of fluctuations. For this reason, the detailed analysis of the initial conditions from inflation carried-out in this work was needed to decide whether the model is ruled out or not.

We have also analysed the vector (helicity 1) and scalar sectors. Also vector  perturbations are generated during inflation leading to a scale invariant vector spectrum with an amplitude which is boosted by a factor $r_I$ with respect to the tensor spectrum. Requiring vector fluctuations to remain perturbative gives an upper limit to the scale of inflation, $H_I < 10^{-2}$~GeV which translates to an inflationary energy scale $V_I^{1/4}< 10^8$~GeV.
In principle, even if the scale of inflation is lower, these vector mode will source tensor modes in the $f$-metric at second order which then can feed into the growing mode. Anyway, a detailed calculation of this is beyond the scope of the present work.

In the  scalar sector we have not discussed the inflaton perturbations, assuming that they are not modified due to the very weak coupling of bigravity during inflation. However, in a bigravity theory we have the helicity-0 mode of the graviton as a second scalar mode. As the Higuchi bound in the scalar sector is violated this mode is growing on super Hubble scales during inflation. We have computed its spectrum at the end of inflation and have found that it is very red, $\propto k^{-4}$.  Requesting that these fluctuations remain perturbative also on the largest scales $k\sim H_0$ gives stringent constraints on the scale of inflation, $H_{I} < 10^{-11}$~GeV, which translates to an inflationary energy scale $V_I^{1/4} \, \lsim \,  10^4$~GeV. 

We conclude that the $\beta_1\beta_4$ model studied in this paper cannot be ruled out  from the analysis of the tensor sector alone. Nevertheless, it is strongly constraint due to the large vector perturbations which are generated during inflation and due to the very red spectrum of scalar perturbations. However, considering that  all the problematic scales of inflation have $H_I >10^{-11}$~GeV, which is far higher than the strong coupling scale, $\La_3$, these results have to be taken with a grain of salt. Still, we recall that this strong coupling limit is derived in a Minkowski background and it is not clear that it should invalidate quantum perturbations on classical Friedmann solutions with a Hubble scale which is larger than $\La_3$.

All other models of bigravity where matter only couples to one of the metrics (the $g$ metric in this work) suffer from exponential instabilities of scalar perturbations on a FLRW background. This makes these models less attractive as candidate solutions to the dark energy problem. Due to the breakdown of linearity one has to work out the theory at higher orders and hope to cure the instabilities, possibly through the Vainshtein mechanism~\cite{Vainshtein:1972sx}. A possible way out is to push the gradient instability to very early times, rendering it unobservable. This can be achieved by lowering the value of the Planck mass of the metric which does not couple to matter \cite{Akrami:2015qga}. In addition, there remain a multitude of massive (bigravity) models whose cosmology deserves further investigation, e.g., where matter, or even different matter sectors, can couple to both metrics \cite{deRham:2014naa,deRham:2014fha,Heisenberg:2014rka,Noller:2014sta}. Alternatively one could also consider non-FLRW backgrounds or change the status of the parameters of the theory,  e.g. by promoting the $\beta_i$ coefficients to functions of the helicity-$0$ mode \cite{deRham:2014gla}, or of some independent scalar field.

\subsection*{Acknowledgments}
We thank Julian Adamek, Yashar Akrami, Luca Amendola, Daniel Figueroa, Matthew Johnson, Frank Koennig, Macarena Lagos, Adam Solomon and Alexandra Terrana for discussions and comments.
This work is supported by the Swiss National Science Foundation.
\newpage
 
 \appendix
 
 \section{Hassan-Rosen bigravity model: general aspects} \label{general setting}

The conventions used for the bigravity action are those of \cite{Hassan:2012wr}. Only one of the two metrics is coupled with matter,  and we restrict to minimal couplings. The action is given by
\be\label{startingaction}
S =   - \int d^4x\sqrt{-g}\left[\frac{M_g^2}{2}(R(g) -2 m^2V(g,f)) +\LL_m(g,\Phi)\right] - \int d^4x\sqrt{-f}\frac{M_f^2}{2}R(f)  \, ,
\ee
where $g$ is the physical metric (the one coupled to matter), $f$ is the second metric, and $M_g=1/\sqrt{8\pi G}\equiv M_p$ and $M_f$ are the respective Planck masses with dimensionless ratio $\al=M_f/M_g$. We assume the matter fields $\Phi$ to be coupled to $g$ only. The potential is given in terms of the tensor field $\mathbb{X} = \sqrt{g^{-1}f}$:
\be
V(g,f) = \sum_{n=0}^{4}\beta_{n}e_{n}(\mathbb{X})\,,
\ee
where the coefficients $\beta_n$ are constants and the polynomials $e_{n}(X)$ are
\begin{eqnarray}
e_{0} &=& \mathbb{I},\, \quad e_{1}=[\mathbb{X}],\, \\ e_{2} &=&\frac{1}{2}([\mathbb{X}]^{2}-[\mathbb{X}^{2}]),\\ 
e_{3} &=& \frac{1}{6}([\mathbb{X}]^{3}-3[\mathbb{X}][\mathbb{X}^{2}]+2[\mathbb{X}^{3}],\\
e_{4} &=& \frac{1}{24}([\mathbb{X}]^{4}-6[\mathbb{X}]^{2}[\mathbb{X}^{2}]+8[\mathbb{X}][\mathbb{X}^{3}] +3[\mathbb{X}^{2}]^{2}-6[\mathbb{X}^{4}]) = \det \mathbb{X} \,.
\end{eqnarray}
The square bracket $\left[\cdots\right]$ denotes the trace.
The equations of motions for $g_{\mu\nu}$ and $f_{\mu\nu}$ are 
\be
R_{\mu\nu}-\frac{1}{2}g_{\mu\nu}\,R+\frac{m^2}{2}\, \sum_{n=0}^3(-)^n\, \beta_n\, \left[g_{\mu\lambda}\, Y_{(n)\nu}^{\lambda}\left(\sqrt{g^{-1}f}\right)+g_{\nu\lambda}\, Y_{(n)\mu}^{\lambda}\left(\sqrt{g^{-1}f}\right)\right]=\frac{1}{M_g^2}\, T_{\mu\nu}\,,
\ee
\be
\bar{R}_{\mu\nu}-\frac{1}{2}f_{\mu\nu}\,\bar{R}+\frac{m^2}{2 \al^{2}}\, \sum_{n=0}^3(-)^n\, \beta_{4-n}\,\left[ f_{\mu\lambda}\, Y_{(n)\nu}^{\lambda}\left(\sqrt{f^{-1}g}\right)+f_{\nu\lambda}\, Y_{(n)\mu}^{\lambda}\left(\sqrt{f^{-1}g}\right)\right]=0\,,
\ee
where the overbar indicates $f_{\mu\nu}$ curvature. The definition of the $Y_{(n)\mu}^{\nu}\left(\mathbb{X}\right)$ matrices is as follows:
\begin{align}
&Y_{(0)}\left(\mathbb{X}\right)=\mathbb{I}\,,\hspace{0.5 cm} Y_{(1)}\left(\mathbb{X}\right)=\mathbb{X} -\mathbb{I}\left[\mathbb{X}\right]\,,\\
&Y_{(2)}\left(\mathbb{X}\right)=\mathbb{X}^2 -\mathbb{X}\left[\mathbb{X}\right]+\frac{1}{2}\mathbb{I}\left(\left[\mathbb{X}\right]^2-\left[\mathbb{X}^2\right]\right)\,,\\
&Y_{(3)}\left(\mathbb{X}\right)=\mathbb{X}^3 -\mathbb{X}^2\left[\mathbb{X}\right]+\frac{1}{2}\mathbb{X}\left(\left[\mathbb{X}\right]^2-\left[\mathbb{X}^2\right]\right)-\frac{1}{6}\mathbb{I}\left(\left[\mathbb{X}\right]^3-3\left[\mathbb{X}\right]\left[\mathbb{X}^2\right]+2\left[\mathbb{X}^3\right]\right)\, .
\end{align}

As a consequence of the Bianchi identity and of the covariant conservation of $T_{\mu\nu}$, we find the following Bianchi constraints (for each one of the two metrics)
\be
\nabla_{\mu}\sum_{n=0}^3(-)^n\, \beta_n\, \left[g_{\mu\lambda}\, Y_{(n)\nu}^{\lambda}\left(\sqrt{g^{-1}f}\right)+g_{\nu\lambda}\, Y_{(n)\mu}^{\lambda}\left(\sqrt{g^{-1}f}\right)\right]=0\,,
\ee
\be
\overline{\nabla}^{\mu}\sum_{n=0}^3(-)^n\, \beta_{4-n}\,\left[ f_{\mu\lambda}\, Y_{(n)\nu}^{\lambda}\left(\sqrt{f^{-1}g}\right)+f_{\nu\lambda}\, Y_{(n)\mu}^{\lambda}\left(\sqrt{f^{-1}g}\right)\right]=0\,,
\ee
where the overbar indicates covariant derivatives with respect to the $f$ metric. Both these constraints follow from the invariance of the interaction term under the diagonal subgroup of the general coordinate transformations of the two metrics.  They are equivalent and in this work we focus on the first one. 

\newpage

\bibliographystyle{utphys}
\bibliography{bi-grefs}

\providecommand{\href}[2]{#2}\begingroup\raggedright\begin{thebibliography}{10}

\bibitem{Lagos:2014lca}
M.~Lagos and P.~G. Ferreira, ``{Cosmological perturbations in massive
  bigravity},'' \href{http://xxx.lanl.gov/abs/1410.0207}{{\tt 1410.0207}}.

\bibitem{Cusin:2014psa}
G.~Cusin, R.~Durrer, P.~Guarato, and M.~Motta, ``{Gravitational waves in
  bigravity cosmology},'' \href{http://xxx.lanl.gov/abs/1412.5979}{{\tt
  1412.5979}}.

\bibitem{Boulware:1973my}
D.~Boulware and S.~Deser, ``{Can gravitation have a finite range?},'' {\em
  Phys.Rev.} {\bf D6} (1972) 3368--3382.

\bibitem{Deser:1967zzb}
S.~Deser, ``{Covariant Decomposition and the Gravitational Cauchy Problem},''
  {\em Ann.Inst.Henri Poincare} {\bf 7} (1967) 149.

\bibitem{deRham:2010ik}
C.~de~Rham and G.~Gabadadze, ``{Generalization of the Fierz-Pauli Action},''
  {\em Phys.Rev.} {\bf D82} (2010) 044020,
  \href{http://xxx.lanl.gov/abs/arXiv:1007.0443}{{\tt arXiv:1007.0443}}.

\bibitem{deRham:2010kj}
C.~de~Rham, G.~Gabadadze, and A.~J. Tolley, ``{Resummation of Massive
  Gravity},'' {\em Phys.Rev.Lett.} {\bf 106} (2011) 231101,
  \href{http://xxx.lanl.gov/abs/1011.1232}{{\tt 1011.1232}}.

\bibitem{Hassan:2011hr}
S.~Hassan and R.~A. Rosen, ``{Resolving the Ghost Problem in non-Linear Massive
  Gravity},'' {\em Phys.Rev.Lett.} {\bf 108} (2012) 041101,
  \href{http://xxx.lanl.gov/abs/1106.3344}{{\tt 1106.3344}}.

\bibitem{Hassan:2011vm}
S.~Hassan and R.~A. Rosen, ``{On Non-Linear Actions for Massive Gravity},''
  {\em JHEP} {\bf 1107} (2011) 009,
  \href{http://xxx.lanl.gov/abs/1103.6055}{{\tt 1103.6055}}.

\bibitem{Hassan:2011tf}
S.~Hassan, R.~A. Rosen, and A.~Schmidt-May, ``{Ghost-free Massive Gravity with
  a General Reference Metric},'' {\em JHEP} {\bf 1202} (2012) 026,
  \href{http://xxx.lanl.gov/abs/1109.3230}{{\tt 1109.3230}}.

\bibitem{Koyama:2011yg}
K.~Koyama, G.~Niz, and G.~Tasinato, ``{Strong interactions and exact solutions
  in non-linear massive gravity},'' {\em Phys.Rev.} {\bf D84} (2011) 064033,
  \href{http://xxx.lanl.gov/abs/1104.2143}{{\tt 1104.2143}}.

\bibitem{Guarato:2013gba}
P.~Guarato and R.~Durrer, ``{Perturbations for massive gravity theories},''
  {\em Phys.Rev.} {\bf D89} (2014) 084016,
  \href{http://xxx.lanl.gov/abs/1309.2245}{{\tt 1309.2245}}.

\bibitem{Gumrukcuoglu:2011zh}
A.~E. Gumrukcuoglu, C.~Lin, and S.~Mukohyama, ``{Cosmological perturbations of
  self-accelerating universe in nonlinear massive gravity},'' {\em JCAP} {\bf
  1203} (2012) 006, \href{http://xxx.lanl.gov/abs/1111.4107}{{\tt 1111.4107}}.

\bibitem{D'Amico:2011jj}
G.~D'Amico, C.~de~Rham, S.~Dubovsky, G.~Gabadadze, D.~Pirtskhalava, and A.~J.
  Tolley, ``{Massive Cosmologies},'' {\em Phys.Rev.} {\bf D84} (2011) 124046,
  \href{http://xxx.lanl.gov/abs/1108.5231}{{\tt 1108.5231}}.

\bibitem{Langlois:2012hk}
D.~Langlois and A.~Naruko, ``{Cosmological solutions of massive gravity on de
  Sitter},'' {\em Class.Quant.Grav.} {\bf 29} (2012) 202001,
  \href{http://xxx.lanl.gov/abs/arXiv:1206.6810}{{\tt arXiv:1206.6810}}.

\bibitem{Fasiello:2012rw}
M.~Fasiello and A.~J. Tolley, ``{Cosmological perturbations in Massive Gravity
  and the Higuchi bound},'' {\em JCAP} {\bf 1211} (2012) 035,
  \href{http://xxx.lanl.gov/abs/arXiv:1206.3852}{{\tt arXiv:1206.3852}}.

\bibitem{Solomon:2014iwa}
A.~R. Solomon, J.~Enander, Y.~Akrami, T.~S. Koivisto, F.~Könnig, and
  E.~Mörtsell, ``{Cosmological viability of massive gravity with generalized
  matter coupling},'' {\em JCAP} {\bf 1504} (2015), no.~04 027,
  \href{http://xxx.lanl.gov/abs/1409.8300}{{\tt 1409.8300}}.

\bibitem{deRham:2014zqa}
C.~de~Rham, ``{Massive Gravity},''
  \href{http://xxx.lanl.gov/abs/1401.4173}{{\tt 1401.4173}}.

\bibitem{Hassan:2011zd}
S.~Hassan and R.~A. Rosen, ``{Bimetric Gravity from Ghost-free Massive
  Gravity},'' {\em JHEP} {\bf 1202} (2012) 126,
  \href{http://xxx.lanl.gov/abs/1109.3515}{{\tt 1109.3515}}.

\bibitem{Hassan:2011ea}
S.~Hassan and R.~A. Rosen, ``{Confirmation of the Secondary Constraint and
  Absence of Ghost in Massive Gravity and Bimetric Gravity},'' {\em JHEP} {\bf
  1204} (2012) 123, \href{http://xxx.lanl.gov/abs/1111.2070}{{\tt 1111.2070}}.

\bibitem{Hassan:2012wr}
S.~Hassan, A.~Schmidt-May, and M.~von Strauss, ``{On Consistent Theories of
  Massive Spin-2 Fields Coupled to Gravity},'' {\em JHEP} {\bf 1305} (2013)
  086, \href{http://xxx.lanl.gov/abs/1208.1515}{{\tt 1208.1515}}.

\bibitem{Akrami:2014lja}
Y.~Akrami, T.~S. Koivisto, and A.~R. Solomon, ``{The nature of spacetime in
  bigravity: two metrics or none?},'' {\em Gen.Rel.Grav.} {\bf 47} (2015),
  no.~1 1838, \href{http://xxx.lanl.gov/abs/1404.0006}{{\tt 1404.0006}}.

\bibitem{Hassan:2014vja}
S.~Hassan, A.~Schmidt-May, and M.~von Strauss, ``{Particular Solutions in
  Bimetric Theory and Their Implications},''
  \href{http://xxx.lanl.gov/abs/1407.2772}{{\tt 1407.2772}}.

\bibitem{deRham:2014fha}
C.~de~Rham, L.~Heisenberg, and R.~H. Ribeiro, ``{Ghosts and matter couplings in
  massive gravity, bigravity and multigravity},'' {\em Phys.Rev.} {\bf D90}
  (2014), no.~12 124042, \href{http://xxx.lanl.gov/abs/1409.3834}{{\tt
  1409.3834}}.

\bibitem{Cusin:2014zoa}
G.~Cusin, J.~Fumagalli, and M.~Maggiore, ``{Non-local formulation of ghost-free
  bigravity theory},'' {\em JHEP} {\bf 1409} (2014) 181,
  \href{http://xxx.lanl.gov/abs/1407.5580}{{\tt 1407.5580}}.

\bibitem{Noller:2014sta}
J.~Noller and S.~Melville, ``{The coupling to matter in Massive, Bi- and
  Multi-Gravity},'' \href{http://xxx.lanl.gov/abs/1408.5131}{{\tt 1408.5131}}.

\bibitem{Akrami:2013ffa}
Y.~Akrami, T.~S. Koivisto, D.~F. Mota, and M.~Sandstad, ``{Bimetric gravity
  doubly coupled to matter: theory and cosmological implications},'' {\em JCAP}
  {\bf 1310} (2013) 046, \href{http://xxx.lanl.gov/abs/1306.0004}{{\tt
  1306.0004}}.

\bibitem{Comelli:2013tja}
D.~Comelli, F.~Nesti, and L.~Pilo, ``{Cosmology in General Massive Gravity
  Theories},'' {\em JCAP} {\bf 1405} (2014) 036,
  \href{http://xxx.lanl.gov/abs/1307.8329}{{\tt 1307.8329}}.

\bibitem{deRham:2014gla}
C.~de~Rham, M.~Fasiello, and A.~J. Tolley, ``{Stable FLRW solutions in
  Generalized Massive Gravity},'' \href{http://xxx.lanl.gov/abs/1410.0960}{{\tt
  1410.0960}}.

\bibitem{Volkov:2011an}
M.~S. Volkov, ``{Cosmological solutions with massive gravitons in the bigravity
  theory},'' {\em JHEP} {\bf 1201} (2012) 035,
  \href{http://xxx.lanl.gov/abs/1110.6153}{{\tt 1110.6153}}.

\bibitem{Comelli:2011zm}
D.~Comelli, M.~Crisostomi, F.~Nesti, and L.~Pilo, ``{FRW Cosmology in Ghost
  Free Massive Gravity},'' {\em JHEP} {\bf 1203} (2012) 067,
  \href{http://xxx.lanl.gov/abs/1111.1983}{{\tt 1111.1983}}.

\bibitem{Konnig:2014dna}
F.~Koennig and L.~Amendola, ``{A minimal bimetric gravity model that fits
  cosmological observations},'' \href{http://xxx.lanl.gov/abs/1402.1988}{{\tt
  1402.1988}}.

\bibitem{Tamanini:2013xia}
N.~Tamanini, E.~N. Saridakis, and T.~S. Koivisto, ``{The Cosmology of
  Interacting Spin-2 Fields},'' {\em JCAP} {\bf 1402} (2014) 015,
  \href{http://xxx.lanl.gov/abs/1307.5984}{{\tt 1307.5984}}.

\bibitem{Fasiello:2013woa}
M.~Fasiello and A.~J. Tolley, ``{Cosmological Stability Bound in Massive
  Gravity and Bigravity},'' {\em JCAP} {\bf 1312} (2013) 002,
  \href{http://xxx.lanl.gov/abs/1308.1647}{{\tt 1308.1647}}.

\bibitem{Solomon:2014dua}
A.~R. Solomon, Y.~Akrami, and T.~S. Koivisto, ``{Cosmological perturbations in
  massive bigravity: I. Linear growth of structures},''
  \href{http://xxx.lanl.gov/abs/1404.4061}{{\tt 1404.4061}}.

\bibitem{vonStrauss:2011mq}
M.~von Strauss, A.~Schmidt-May, J.~Enander, E.~Mortsell, and S.~Hassan,
  ``{Cosmological Solutions in Bimetric Gravity and their Observational
  Tests},'' {\em JCAP} {\bf 1203} (2012) 042,
  \href{http://xxx.lanl.gov/abs/1111.1655}{{\tt 1111.1655}}.

\bibitem{Berg:2012kn}
M.~Berg, I.~Buchberger, J.~Enander, E.~Mortsell, and S.~Sjors, ``{Growth
  Histories in Bimetric Massive Gravity},''
  \href{http://xxx.lanl.gov/abs/1206.3496}{{\tt 1206.3496}}.

\bibitem{2013JHEP...03..099A}
Y.~{Akrami}, T.~S. {Koivisto}, and M.~{Sandstad}, ``{Accelerated expansion from
  ghost-free bigravity: a statistical analysis with improved generality},''
  {\em Journal of High Energy Physics} {\bf 3} (Mar., 2013) 99,
  \href{http://xxx.lanl.gov/abs/1209.0457}{{\tt 1209.0457}}.

\bibitem{Konnig:2013gxa}
F.~Koennig, A.~Patil, and L.~Amendola, ``{Viable cosmological solutions in
  massive bimetric gravity},'' {\em JCAP} {\bf 1403} (2014) 029,
  \href{http://xxx.lanl.gov/abs/1312.3208}{{\tt 1312.3208}}.

\bibitem{Akrami:2015qga}
Y.~Akrami, S.~Hassan, F.~Könnig, A.~Schmidt-May, and A.~R. Solomon,
  ``{Bimetric gravity is cosmologically viable},''
  \href{http://xxx.lanl.gov/abs/1503.07521}{{\tt 1503.07521}}.

\bibitem{Konnig:2015lfa}
F.~Koennig, ``{On Higuchi Ghosts and Gradient Instabilities in Bimetric
  Gravity},'' \href{http://xxx.lanl.gov/abs/1503.07436}{{\tt 1503.07436}}.

\bibitem{Gumrukcuoglu:2015nua}
A.~E. Gumrukcuoglu, L.~Heisenberg, S.~Mukohyama, and N.~Tanahashi, ``{Cosmology
  in bimetric theory with an effective composite coupling to matter},''
  \href{http://xxx.lanl.gov/abs/1501.02790}{{\tt 1501.02790}}.

\bibitem{Comelli:2015pua}
D.~Comelli, M.~Crisostomi, K.~Koyama, L.~Pilo, and G.~Tasinato, ``{Cosmology of
  bigravity with doubly coupled matter},''
  \href{http://xxx.lanl.gov/abs/1501.00864}{{\tt 1501.00864}}.

\bibitem{Enander:2014xga}
J.~Enander, A.~R. Solomon, Y.~Akrami, and E.~Mortsell, ``{Cosmic expansion
  histories in massive bigravity with symmetric matter coupling},'' {\em JCAP}
  {\bf 01} (2015) 006, \href{http://xxx.lanl.gov/abs/1409.2860}{{\tt
  1409.2860}}.

\bibitem{Comelli:2012db}
D.~Comelli, M.~Crisostomi, and L.~Pilo, ``{Perturbations in Massive Gravity
  Cosmology},'' {\em JHEP} {\bf 1206} (2012) 085,
  \href{http://xxx.lanl.gov/abs/1202.1986}{{\tt 1202.1986}}.

\bibitem{Khosravi:2012rk}
N.~Khosravi, H.~R. Sepangi, and S.~Shahidi, ``{Massive cosmological scalar
  perturbations},'' {\em Phys. Rev.} {\bf D86} (2012) 043517,
  \href{http://xxx.lanl.gov/abs/1202.2767}{{\tt 1202.2767}}.

\bibitem{Comelli:2014bqa}
D.~Comelli, M.~Crisostomi, and L.~Pilo, ``{FRW Cosmological Perturbations in
  Massive Bigravity},'' {\em Phys.Rev.} {\bf D90} (2014), no.~8 084003,
  \href{http://xxx.lanl.gov/abs/1403.5679}{{\tt 1403.5679}}.

\bibitem{DeFelice:2014nja}
A.~De~Felice, A.~E. Gumrukcuoglu, S.~Mukohyama, N.~Tanahashi, and T.~Tanaka,
  ``{Viable cosmology in bimetric theory},'' {\em JCAP} {\bf 1406} (2014) 037,
  \href{http://xxx.lanl.gov/abs/1404.0008}{{\tt 1404.0008}}.

\bibitem{Kuhnel:2012gh}
F.~Kuhnel, ``{Instability of certain bimetric and massive-gravity theories},''
  {\em Phys.Rev.} {\bf D88} (2013), no.~6 064024,
  \href{http://xxx.lanl.gov/abs/1208.1764}{{\tt 1208.1764}}.

\bibitem{Amendola_pert}
F.~Koennig, Y.~Akrami, L.~Amendola, M.~Motta, and A.~R. Solomon, ``{Stable and
  unstable cosmological models in bimetric massive gravity},''
  \href{http://xxx.lanl.gov/abs/1407.4331}{{\tt 1407.4331}}.

\bibitem{Amendola:2015tua}
L.~Amendola, F.~Koennig, M.~Martinelli, V.~Pettorino, and M.~Zumalacarregui,
  ``{Surfing gravitational waves: can bigravity survive growing tensor
  modes?},'' \href{http://xxx.lanl.gov/abs/1503.02490}{{\tt 1503.02490}}.

\bibitem{Fasiello:2015csa}
M.~Fasiello and R.~H. Ribeiro, ``{Mild bounds on bigravity from primordial
  gravitational waves},'' \href{http://xxx.lanl.gov/abs/1505.00404}{{\tt
  1505.00404}}.

\bibitem{Higuchi:1986py}
A.~Higuchi, ``{Forbidden mass range for spin-2 field theory in De Sitter
  space-time},'' {\em Nucl.Phys.} {\bf B282} (1987) 397.

\bibitem{Johnson:2015tfa}
M.~Johnson and A.~Terrana, ``{Tensor Modes in Bigravity: Primordial to
  Present},'' \href{http://xxx.lanl.gov/abs/1503.05560}{{\tt 1503.05560}}.

\bibitem{ArkaniHamed:2002sp}
N.~Arkani-Hamed, H.~Georgi, and M.~D. Schwartz, ``{Effective field theory for
  massive gravitons and gravity in theory space},'' {\em Annals Phys.} {\bf
  305} (2003) 96--118, \href{http://xxx.lanl.gov/abs/hep-th/0210184}{{\tt
  hep-th/0210184}}.

\bibitem{Abramo}
M.~Abramowitz and I.~Stegun, {\em Handbook of Mathematical Functions}.
\newblock Dover Publications, New York, 1972.

\bibitem{Ade:2015xua}
{\bf Planck} Collaboration, P.~A.~R. Ade {\em et.~al.}, ``{Planck 2015 results.
  XIII. Cosmological parameters},''
  \href{http://xxx.lanl.gov/abs/1502.01589}{{\tt 1502.01589}}.

\bibitem{German:2001tz}
G.~German, G.~G. Ross, and S.~Sarkar, ``{Low scale inflation},'' {\em
  Nucl.Phys.} {\bf B608} (2001) 423--450,
  \href{http://xxx.lanl.gov/abs/hep-ph/0103243}{{\tt hep-ph/0103243}}.

\bibitem{Allahverdi:2006iq}
R.~Allahverdi, K.~Enqvist, J.~Garcia-Bellido, and A.~Mazumdar, ``{Gauge
  invariant MSSM inflaton},'' {\em Phys.Rev.Lett.} {\bf 97} (2006) 191304,
  \href{http://xxx.lanl.gov/abs/hep-ph/0605035}{{\tt hep-ph/0605035}}.

\bibitem{Bonvin:2011dt}
C.~Bonvin, C.~Caprini, and R.~Durrer, ``{Magnetic fields from inflation: the
  transition to the radiation era},'' {\em Phys.Rev.} {\bf D86} (2012) 023519,
  \href{http://xxx.lanl.gov/abs/1112.3901}{{\tt 1112.3901}}.

\bibitem{Vainshtein:1972sx}
A.~Vainshtein, ``{To the problem of nonvanishing gravitation mass},'' {\em
  Phys.Lett.} {\bf B39} (1972) 393--394.

\bibitem{deRham:2014naa}
C.~de~Rham, L.~Heisenberg, and R.~H. Ribeiro, ``{On couplings to matter in
  massive (bi-)gravity},'' {\em Class. Quant. Grav.} {\bf 32} (2015) 035022,
  \href{http://xxx.lanl.gov/abs/1408.1678}{{\tt 1408.1678}}.

\bibitem{Heisenberg:2014rka}
L.~Heisenberg, ``{Quantum corrections in massive bigravity and new effective
  composite metrics},'' {\em Class. Quant. Grav.} {\bf 32} (2015), no.~10
  105011, \href{http://xxx.lanl.gov/abs/1410.4239}{{\tt 1410.4239}}.

\end{thebibliography}\endgroup


\providecommand{\href}[2]{#2}\begingroup\raggedright\endgroup

\end{document}